\documentclass[sigconf, nonacm]{acmart}

\newcommand\vldbdoi{XX.XX/XXX.XX}
\newcommand\vldbpages{XXX-XXX}
\newcommand\vldbvolume{15}
\newcommand\vldbissue{1}
\newcommand\vldbyear{2022}
\newcommand\vldbauthors{\authors}
\newcommand\vldbtitle{\shorttitle} 
\newcommand\vldbavailabilityurl{}
\newcommand\vldbpagestyle{plain} 

\usepackage{graphicx}
\usepackage{balance}  
\usepackage{comment}
\usepackage{xspace}
\usepackage{cite}
\usepackage{enumitem}
\usepackage{algorithm}
\usepackage[noend]{algpseudocode}
\usepackage{multirow}
\usepackage{fancyvrb}
\usepackage{listings}

\usepackage{colortbl}
\usepackage{hhline}
\usepackage{listings}
\usepackage{newfloat}
\usepackage{fancyvrb}

\definecolor{mahogany}{HTML}{85200c}

\newcommand{\red}[1]{{\color{black} {#1}}}
\newcommand{\dred}[1]{{\color{mahogany} {#1}}}

\newcommand{\sys}{\textsc{Ember}\xspace}
\newcommand{\record}{\texttt{record}\xspace}
\newcommand{\sentence}{\texttt{sentence}\xspace}
\newcommand{\encoder}{\texttt{encoder}\xspace}
\newcommand{\labeler}{\texttt{labeler}\xspace}
\newcommand{\embedding}{\texttt{embedding}\xspace}
\newcommand{\loss}{\texttt{loss}\xspace}
\newcommand{\sampler}{\texttt{sampler}\xspace}
\newcommand{\preparer}{\texttt{preparer}\xspace}
\newcommand{\retriever}{\texttt{retriever}\xspace}
\newcommand{\records}{\texttt{record}s\xspace}
\newcommand{\sentences}{\texttt{sentence}s\xspace}
\newcommand{\encoders}{\texttt{encoder}s\xspace}
\newcommand{\labelers}{\texttt{labeler}s\xspace}
\newcommand{\embeddings}{\texttt{embedding}s\xspace}
\newcommand{\losses}{\texttt{loss}es\xspace}
\newcommand{\samplers}{\texttt{sampler}s\xspace}
\newcommand{\preparers}{\texttt{preparer}s\xspace}
\newcommand{\retrievers}{\texttt{retriever}s\xspace}

\newcommand{\minihead}[1]{{\vspace{.45em}\noindent\textbf{#1.} }}

\newcommand{\eat}[1]{}

\newcommand*{\vsepfbox}[1]{%
  \begingroup
    \sbox0{\fbox{#1}}%
    \setlength{\fboxrule}{0pt}%
    \mbox{\kern-\fboxsep\fbox{\unhbox0}\kern-\fboxsep}%
  \endgroup
}

\DeclareFloatingEnvironment[
  fileext = los ,
  listname = {List of listings} ,
  name = Listing
]{llisting}

\begin{document}
\title{\sys: No-Code Context Enrichment via \\ Similarity-Based Keyless Joins}

\eat{\author{Ben Trovato}
\affiliation{%
  \institution{Institute for Clarity in Documentation}
  \streetaddress{P.O. Box 1212}
  \city{Dublin}
  \state{Ireland}
  \postcode{43017-6221}
}
\email{trovato@corporation.com}
}

\author{Sahaana Suri, Ihab F. Ilyas$^\ast$, Christopher R\'e, Theodoros Rekatsinas$^\dagger$}
\affiliation{%
  \institution{Stanford University, University of Waterloo$^\ast$, UW-Madison$^\dagger$}
}

\begin{abstract}
Structured data, or data that adheres to a pre-defined schema, can suffer from fragmented context: information describing a single entity can be scattered across multiple datasets or tables tailored for specific business needs, with no explicit linking keys (e.g., primary key-foreign key relationships or heuristic functions).
Context enrichment, or rebuilding fragmented context, using \emph{keyless joins} is an implicit or explicit step in machine learning (ML) pipelines over structured data sources.
This process is tedious, domain-specific, and lacks support in now-prevalent no-code ML systems that let users create ML pipelines using just input data and high-level configuration files.
In response, we propose \sys, a system that abstracts and automates keyless joins to generalize context enrichment.
Our key insight is that \sys can enable a general keyless join operator by constructing an index populated with task-specific embeddings. 
\sys learns these embeddings by leveraging Transformer-based representation learning techniques.
We describe our core architectural principles and operators when developing \sys, and empirically demonstrate that \sys allows users to develop no-code pipelines for five domains, including search, recommendation and question answering, and can exceed alternatives by up to 39\% recall, with as little as a single line configuration change.

\end{abstract}

\maketitle

\pagestyle{\vldbpagestyle}
\begingroup\small\noindent\raggedright\textbf{PVLDB Reference Format:}\\
\vldbauthors. \vldbtitle. PVLDB, \vldbvolume(\vldbissue): \vldbpages, \vldbyear.\\
\href{https://doi.org/\vldbdoi}{doi:\vldbdoi}
\endgroup
\begingroup
\renewcommand\thefootnote{}\footnote{\noindent
This work is licensed under the Creative Commons BY-NC-ND 4.0 International License. Visit \url{https://creativecommons.org/licenses/by-nc-nd/4.0/} to view a copy of this license. For any use beyond those covered by this license, obtain permission by emailing \href{mailto:info@vldb.org}{info@vldb.org}. Copyright is held by the owner/author(s). Publication rights licensed to the VLDB Endowment. \\
\raggedright Proceedings of the VLDB Endowment, Vol. \vldbvolume, No. \vldbissue\ %
ISSN 2150-8097. \\
\href{https://doi.org/\vldbdoi}{doi:\vldbdoi} \\
}\addtocounter{footnote}{-1}\endgroup

\ifdefempty{\vldbavailabilityurl}{}{
\vspace{.3cm}
\begingroup\small\noindent\raggedright\textbf{PVLDB Artifact Availability:}\\
The source code, data, and/or other artifacts have been made available at \url{\vldbavailabilityurl}.
\endgroup
}

\section{Introduction}

\begin{figure}
\includegraphics[width=\columnwidth]{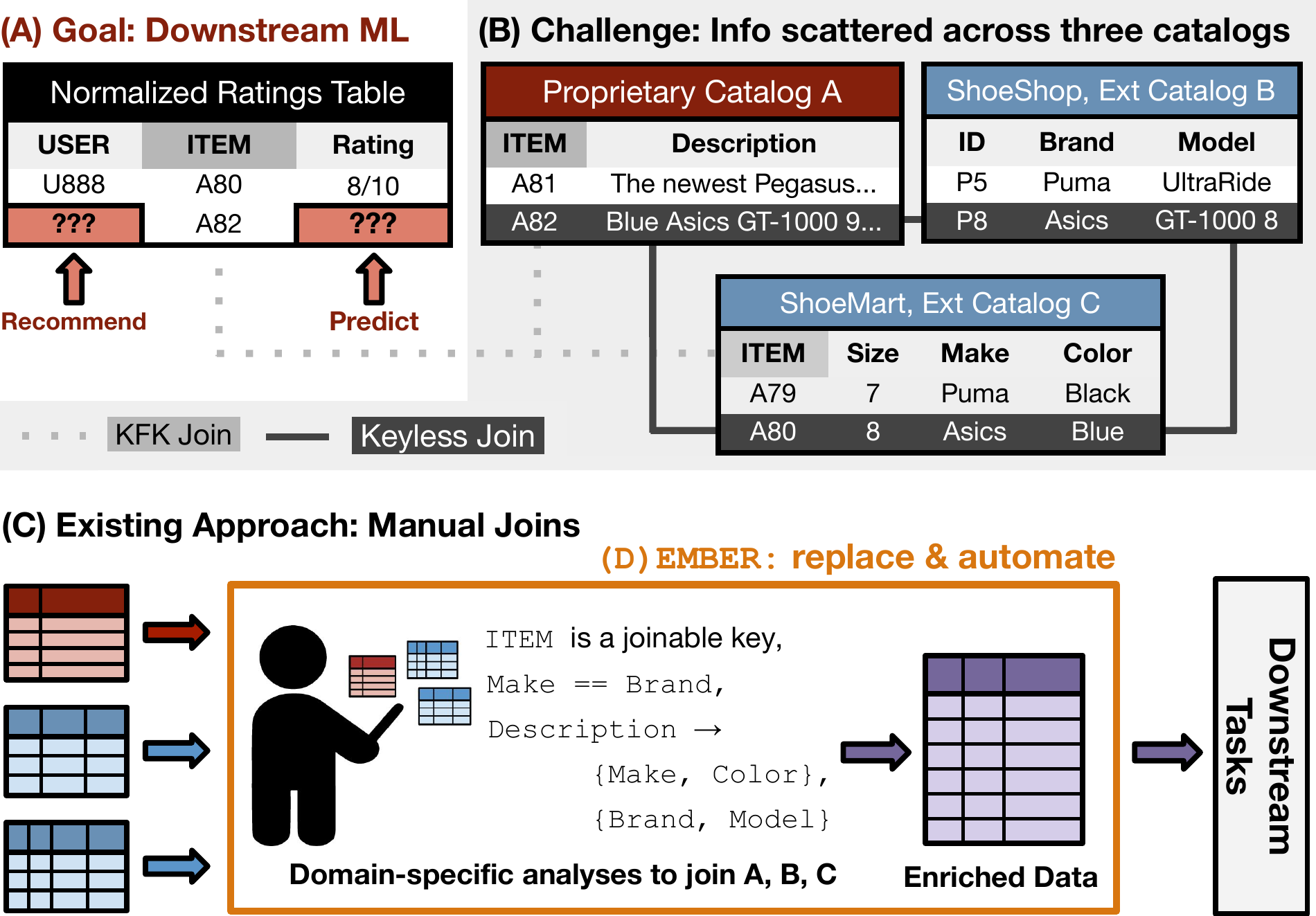}
\caption{An end-to-end task requiring context enrichment. Predicting the rating of and recommending a new product (\texttt{A82}), requires relating the Asics products (highlighted in dark gray) via a keyless join (top). This process is manual due to data heterogeneity---we aim to automate it (bottom).}
\label{fig:example}
\vspace{-1em}
\end{figure}

Machine learning (ML) systems that extract structural and semantic context from unstructured datasets have revolutionzed domains such as  computer vision~\citep{alexnet} and natural language processing~\citep{bert,gpt2}.
Unfortunately, applying these sytems to structured and semi-structured data repositories that consist of datasets with pre-defined schemas is challenging as their context is often \textit{fragmented}: they frequently scatter information regarding a data record across domain-specific datasets with unique schemas.
For instance, in Figure~\ref{fig:example}B, information regarding \textsc{Asics} shoes is scattered across three catalogs with unique schemas.
These datasets adhere to fixed schemas that are optimized for task-specific querying, and often lack explicit linking keys, such as primary key-foreign key (KFK) relationships.
This constrains users to a single view of an entity that is specialized for a specific business need.

Associating these fragmented data contexts is critical in enabling ML-powered applications over such datasets---a process we denote as \emph{context enrichment}---yet is a heavily manual endeavor due to task and dataset heterogeneity.
Engineers develop solutions for context enrichment tailored to their task, such as similarity-based blocking in data integration~\citep{blocking}, retriever models in question answering~\citep{retriever}, or retrieval functions in search~\citep{bm25} (see Section~\ref{sec:applications}). 
Constructing these independent solutions is repetitive, time-consuming, and results in a complicated landscape of overlapping, domain-specific methods.
For instance, consider the scenario depicted in Figure~\ref{fig:example}:

\vspace{-7pt}
\emph{\\An e-commerce company has a proprietary product catalog, and aggregates product information from several external vendors to perform market analysis. 
Each vendor uses a unique product catalog, each with unique product representations (i.e., schema) ranging from free-form text to tabular product descriptions; products may overlap and evolve (Figure~\ref{fig:example}B).
Given normalized tables containing user and rating data, an engineer wishes to estimate the rating for a candidate new product (\texttt{A82}) and identify users to recommend the product to (Figure~\ref{fig:example}A).
}
\vspace{5pt}

The engineer must first perform context enrichment by joining information across tables to extract features that capture similarities between the new (\texttt{A82}) and existing (\texttt{A80}, \texttt{P8}) products.
They can then estimate the product rating, and recommend the new product to users based on how they rated related products.

The classic data management approach is to denormalize datasets using KFK joins. 
This fails to solve the problem due to two reasons.
First, not all tables can be joined when relying on only KFK relationships (e.g., there is no KFK relationship between Catalog B and Catalogs A or C). 
Second, even when KFK relationships exist, as between Catalogs A and C (\textsc{\texttt{ITEM}}), relying on only KFK semantics fails to capture the similarity between \texttt{A82} and \texttt{A80}.

Alternatively, the engineer can rely on similarity-based join techniques, as in data blocking~\citep{blocking}, built to avoid exhaustive, pairwise comparison of potentially joinable records.
However, as we show in Section~\ref{subsec:general}, the optimal choice of join operator to maximize recall of relevant records is task-dependent, and may not scale at query time when new records arrive.
The engineer must first note that the \texttt{Description} column in Catalog A relates to the \texttt{Brand} and \texttt{Model} columns in Catalog B and the \texttt{Size}, \texttt{Make}, and \texttt{Color} columns in Catalog C.
They can then select a custom join based on table properties: for matching primarily short, structured data records, they may want to join based on Jaccard similarity, whereas BM25 may be better suited for purely textual records (see Table~\ref{tab:baselines}).
As database semantics do not natively support these \emph{keyless joins} that require similarity-based indexing, joining arbitrary catalogs remains heavily manual---even large companies rely on vendors to categorize listings, which results in duplicate listings.\footnote{https://sell.amazon.com/sell.html}

To counter this manual process, we draw inspiration from recent no-code ML systems such as Ludwig~\citep{ludwig}, H20.ai~\citep{h20}, and Data Robot~\citep{robot} that are rapidly shifting practitioners towards higher-level configuration-based abstractions for developing ML applications.
Despite their success, these systems leave context enrichment as a user-performed data preparation step.\footnote{https://cloud.google.com/automl-tables/docs/prepare}
In this paper, we evaluate how to bring no-code semantics to context enrichment. 

No out-of-the-box query interface surfaces the relatedness between \texttt{A80}, \texttt{A82}, and \texttt{P8} (all \texttt{Asics}, two \texttt{GT-1000}, and two \texttt{blue}), and links these records to the Ratings table with minimal intervention. 
The challenge in developing a no-code context enrichment system to enable this is to construct an architecture that is simultaneously:
\begin{enumerate}
    \item \textbf{General}: Applicable to a wide variety of tasks and domains.
    \item \textbf{Extensibile}: Customizable for domain or downstream needs.
    \item \textbf{Low Effort}: Usable with minimal configuration.
\end{enumerate}

Our key insight to enable such a system is to simplify context enrichment by abstracting an interface for a new class of join: a \textit{learned keyless join} that operates over record-level similarity. 
Just as traditional database joins provide an abstraction layer for combining structured data sources given KFK relationships, we formalize keyless joins as an abstraction layer for context enrichment. 

We then propose \sys: a no-code context enrichment framework that implements a keyless join abstraction layer.
\sys creates an index populated with task-specific embeddings that can be quickly retrieved at query time, and can operate over arbitrary semi-structured datasets with unique but fixed schema.  
To provide generality, \sys relies on Transformers~\citep{transformers} as building blocks for embedding generation, as they have demonstrated success across textual, semi-structured, and structured workloads~\citep{picket,tabert,bert,gpt2,neuraldb}.
To provide extensibility, \sys is composed of a modular, three step architecture with configurable operators. 
To provide ease of use, \sys can be configured using a simple json-based configuration file, and provides a default configuration that works well across five tasks with a single line change per task.

\begin{figure}
\includegraphics[width=\columnwidth]{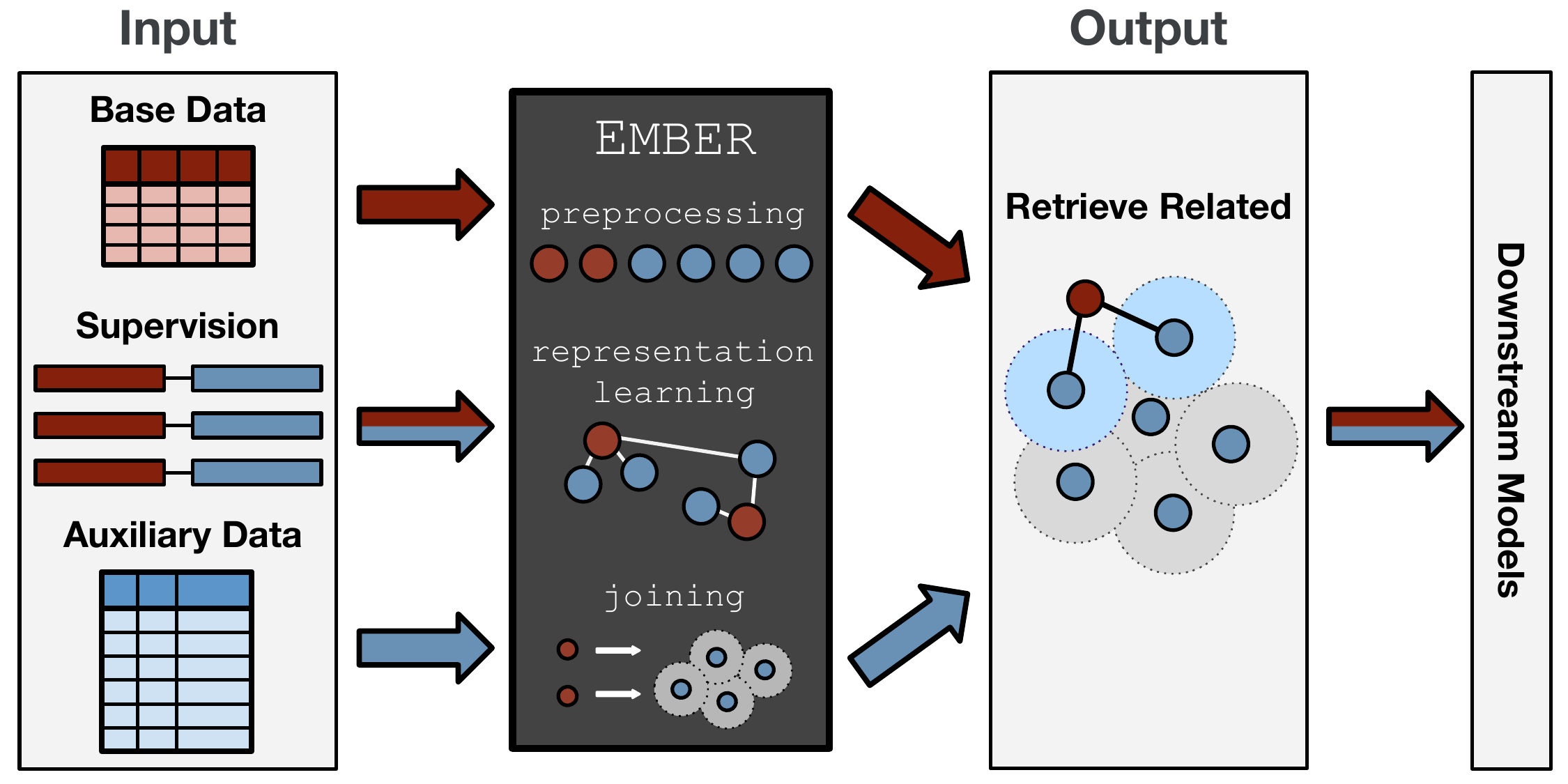}
\caption{\sys's interface for context enrichment. \sys retrieves related auxiliary records for each base record.}
\label{fig:overview}
\vspace{-1em}
\end{figure}

As input, users provide \sys with: \textbf{(1)} a base data source, \textbf{(2)} an auxiliary data source, \textbf{(3)} a set of examples of related records across the sources.
For each record in the base data source, \sys returns related records from the auxiliary data source as characterized by the examples, which can be post-processed for downstream tasks (e.g., concatenated or averaged).
We present \sys's three-step modular architecture that enables index construction via Transformer-based embeddings to facilitate related record retrieval: preprocessing, representation learning, and joining (Figure~\ref{fig:overview}).

\eat{
\begin{enumerate}
    \item A base data source
    \item An auxiliary data source
    \item A set of examples of related records across the sources
\end{enumerate}
}

\minihead{Preprocessing} \sys transforms records from different data sources to a common representation, which allows us to apply the same methods across tasks. 
By default, \sys uses operators that convert each record to a natural language sentence for input to a Transformer-based encoder, which we optionally pretrain using self-supervision.
We demonstrate that depending on the data vocabulary, encoder pretraining to bootstrap the pipeline with domain knowledge can increase recall of relevant records by up to \red{20.5\%}. 

\minihead{Representation Learning} \sys tunes the preprocessed representations to identify contextually similar records as defined by supervised examples.
Learning this representation is task-dependent: we show that a natural approach of using pretrained Transformer-based embeddings with no fine-tuning performs up to three orders of magnitude worse than a fine-tuned approach, often returning less than \red{10\%} of relevant records.  
Pretraining in the first step can increase recall by over \red{30\%}, but this is still insufficient.
Thus, \sys relies on operators that use supervised examples of related records and a contrastive triplet loss~\citep{tripletloss} to learn a representation that encourages similar records to be close while pushing away dissimilar records in the induced embedding space. 
We evaluate how many labeled examples \sys needs to match the performance of using all provided examples, and find that at times, \red{1\%} of the labels is sufficient, and that operators for hard negative sampling improve performance by up to \red{30\%} recall, further improving ease of use.

\minihead{Joining} \sys quickly retrieves related records using the tuned representations to perform the keyless join.
 \sys populates an index with the embeddings learned to capture record similarity, and uses optimized maximum inner product search (MIPS) to identify the nearest neighbors of a given record.
This procedure allows \sys to capture one-to-one and one-to-many join semantics across a variety of downstream applications with low query-time overhead.
We demonstrate that \sys meets or exceeds the performance of eight similarity-based join baselines including BM25~\citep{bm25}, Jaccard-similarity-based joins, Levenshtein distance-based joins, and AutoFuzzyJoin~\citep{autofj} with respect to recall and query runtime.

Although conventional wisdom in data management often says to rely on domain-specific pipelines for context enrichment, we introduce no-code ML semantics to context enrichment and empirically show how a single system can generalize to heterogeneous downstream tasks.
We report on our experiences in deploying \sys across five tasks: fuzzy joining, entity matching, question answering, search, and recommendation.
We demonstrate that \sys generalizes to each, requires as little as a single line configuration change, and extends to downstream similarity-based analyses for search, entity matching, and recommendation as in Figure~\ref{fig:example}. 

In summary, we present the following contributions in this paper:

\begin{itemize}
    \setlength\itemsep{.75em}
    \item We propose keyless joins with a join specification to serve as an abstraction layer for context enrichment. To our knowledge, this is the first work to generalize similarity-based tasks across data integration, natural language processing, search, and recommendation as a data management problem.
    \item We design and develop \sys, the first no-code framework for context enrichment that implements keyless joins, and provides an API for extending and optimizing keyless joins.
    \item We empirically demonstrate that \sys generalizes to five workloads by meeting or exceeding the recall of baselines while using \red{6} or fewer configuration line changes, and evaluate the modular design of \sys's default architecture.
\end{itemize}

\section{Context Enrichment}
\label{sec:applications}

In this section, we define context enrichment, and provide example workloads that can be framed as a form of context enrichment.

\subsection{Problem Statement}
\label{subsec:problem}

Structured data, or data that adheres to a fixed schema formatted in rows and columns, suffers from context fragmentation.
We broadly define structured data to include semi-structured and textual data sources that may be stored and keyed in data management systems, such as content from Wikipedia articles or web search results. 
Unlike unstructured data, structured data follows a scattered format that is efficient to store, query, and manipulate for specific business needs.
This may be in the form of an e-commerce company storing datasets cataloging products and user reviews as in Figure~\ref{fig:example}.
To use this fragmented data for ML tasks, practitioners implicitly or explicitly join these dataset to construct discriminative features. 

We refer to this joining process as \emph{context enrichment}, which we define as follows:
Given a base dataset in tabular form, $D_0$, context enrichment aligns $D_0$ with context derived from auxiliary data sources, $D = \{D_1, ... , D_M\}$, to solve a task $T$. 
We focus on a text-based, two-dataset case ($|D| = 1$), but propose means for multi-dataset extension Sections~\ref{sec:preprocessing}-\ref{sec:joining}, and defer their evaluation to future work.
We explore context enrichment in regimes with abundant and limited labeled relevant pairs $Y$ to learn source alignment.


We represent each dataset $D_i$ as an $\mathbf{n_i}\times\mathbf{d_i}$ matrix of $\mathbf{n_i}$ data points (or records) and  $\mathbf{d_i}$ columns. 
We denote $C_{ir}$ as the $r^{th}$ column in dataset $i$.
We denote $d_{ij}$ as the $j^{th}$ row in dataset $i$.
Columns are numeric or textual, and we refer to the domain of $d_{ij}$ as $\mathbf{D_i}$.

\eat{Each dataset $D_i$ consists of $\mathbf{d_i}$ columns, each of which we denote as $C_{ir}$, where $r \in \{1, ...,\mathbf{d_i}\}$. 
Columns are either numeric or textual, meaning $C_{ir} \in X^{{n_i}}$, where $X \in \{\mathbb{R} \cup \mathbb{N}\}$ and $n_i$ is the number of records in dataset $D_i$.
Data points in each dataset are represented as $d_{ij}$ for the $j^{th}$ row in dataset $i$, where $j \in \{1, ...,n_i\}$ and $d_{ij} \in X^{\mathbf{d_i}}$. We describe how to extend beyond these formats in Section~\ref{sec:representation}. }


Our goal is to \emph{enrich}, or identify all context for, each $d_{0j}$ in $D_0$: auxiliary records $d_{lm}$ ($l \neq 0$) that are related to $d_{0j}$, as per task $T$.

\subsection{Motivating Applications}
\label{subsec:target}

Context enrichment is a key component for applications across a broad array of domains. 
Thus, a context enrichment system would reduce the developer time and effort needed to construct domain-specific pipelines.
We now describe a subset of these domains, which we evaluate in Section~\ref{sec:eval}. 
Additional applicable domains include entity linkage or disambiguation~\citep{link2,link1}, nearest neighbor machine translation~\citep{knnmt}, and nearest neighbor language modeling~\citep{knnlm}.

\minihead{Entity Matching and Deduplication}
Entity matching identifies data points that refer to the same real-world entity across two different collections of entity mentions~\citep{matchsurvey}. 
An entity denotes any distinct real-world object such as a person or organization, while its entity mention is a reference to this entity in a structured dataset or a text span.   
Entity deduplication identifies entities that are found multiple times in a single data source, thus is a special case of entity matching where both collections of entity mentions are the same.

\vspace{4pt}
\noindent
We frame entity matching as a context enrichment problem where $D_0$ and $D_1$ represent the two collections of entity mentions. 
Context enrichment aligns entities $d_{1m}$ in the auxiliary dataset $D_1$ with each entity $d_{0j}$ in the base dataset $D_0$. In entity deduplication, $D_0 = D_1$. 
The downstream task $T$ is a binary classifier over the retrieved relevant records, as the aim is to identify matching entities. 

\eat{
\minihead{Entity Linking or Disambiguation}
Entity linking, or named entity disambiguation, joins entity mentions (as defined above) in the form of a text span to a canonical entity in a database or knowledge graph~\citep{link1,link2}. Entity linkage differs from entity matching in that it has access to relationship information across entities.

\vspace{3pt}
\noindent
We frame entity linkage as a context enrichment problem where $D_0$ and $D_1$ represent the collection of entity mentions and a database or linearized representation of the knowledge graph, respectively. 
Context enrichment aligns entities $d_{1m}$ in the auxiliary dataset $D_1$ that map to each entity $d_{0j}$ in the base dataset $D_0$.
The downstream task $T$ is a binary classifier over the retrieved relevant records, as the aim is to identify matching entities. \\
}

\minihead{Fuzzy Joining}
A fuzzy join identifies data point pairs that are similar to one another across two database tables, where similar records may be identified with respect to a similarity function (e.g., cosine similarity, Jaccard similarity, edit distance) and threshold~\citep{fj1,surajit} defined over a subset of key columns. 
Though related to entity matching, fuzzy joins can be viewed as a primitive used to efficiently mine and block pairs that are similar across the two data sources. 

\vspace{4pt}
\noindent
We frame fuzzy joining as a context enrichment problem where $D_0$ and $D_1$ represent the two database tables.
We let $C_0$ and $C_1$ denote the set of key columns used for joining.
Records $d_{0j}$ and $d_{1m}$ are joined if their values in columns $C_0$ and $C_1$ are similar, to some threshold quantity. 
This is equivalent to a generic context enrichment task with a limited number of features used. 
The downstream task $T$ is a top-k query, as the aim is to identify similar entities. 

\minihead{Recommendation}
A recommender system predicts the rating that a user would give an item, typically for use in content recommendation or filtering~\citep{recsys}. 
We also consider the broader problem of uncovering the global rating for an item, as in the e-commerce example in Figure~\ref{fig:example}, as being a form of recommendation problem---namely, the global score may be predicted by aggregating the predicted per-user results obtained via classic recommender systems.  

\vspace{4pt}
\noindent
We frame such recommendation workloads as a context enrichment problem where $D_0$ represents all information regarding the entity to be ranked (e.g., a base product table, or new candidate products), and $D_1$ represents the information of the rankers, or other auxiliary data (e.g., catalogs of previously ranked products).
Context enrichment aligns rankers $d_{1m}$ in the auxiliary dataset $D_1$ with those that are related to each entity to be ranked $d_{0j}$ in the base dataset $D_0$. 
The downstream task $T$ is to return the top-1 query, or to perform aggregation or train an ML model over the returned top-k entries.

\minihead{Search}
An enterprise search engine returns information relevant to a user query issued over internal databases, documentation, files, or web pages~\citep{enterprise1, enterprise2}. 
A general web retrieval system displays information relevant to a given search query~\citep{search1,search2}. 
Both rely on information retrieval techniques and relevance-based ranking over an underlying document corpus as building blocks to develop complex, personalized pipelines, or for retrospective analyses~\citep{marco}.

\vspace{4pt}
\noindent
We frame retrieval and ranking in search as a context enrichment problem where $D_0$ represents the set of user queries, and $D_1$ represents the underlying document corpus.
Context enrichment aligns documents $d_{1m}$ in the auxiliary dataset $D_1$ that map to each query $d_{0j}$ in the base dataset $D_0$.
The downstream task $T$ is to return the top-k documents for each query, sorted by their query relatedness.

\minihead{Question Answering}
Question answering systems answer natural language questions given a corresponding passage (e.g., in reading comprehension tasks) or existing knowledge source (e.g., in open domain question answering)~\citep{squad,qa}.
Multi-hop question answering generalizes this setting, where the system must traverse a series of passages to uncover the answer~\citep{multihop}. 
For instance, the question "what color is the ocean?" may be provided the statements "the sky is blue" and "the ocean is the same color as the sky." 

\vspace{4pt}
\noindent
We frame the retriever component~\citep{retriever} of a classic retriever-reader model in open domain question answering as a context enrichment problem. 
Our task is to identify the candidate text spans in the provided passages that contain the question's answer (the retriever component), and a downstream reader task can formulate the answer to the question.
We let dataset $D_0$ represent the set of questions, and $D_1$ represent the provided passages or knowledge source, split into spans (e.g., based on sentences, or fixed word windows).
Context enrichment aligns spans $d_{1m}$ in the auxiliary dataset $D_1$ that contain the answer to each question $d_{0j}$ in the base dataset $D_0$. 
Multi-hop question answering requires an extra round of context enrichment for each hop required.
Each round selects a different passage's spans as the base table $D_0$; to identify context relevant to the question, the user can traverse the related spans extracted from each passage.
The downstream task $T$ is to learn a reader model to answer questions using the enriched data sources.

\eat{
\minihead{Machine Translation}

\red{Do the nearest neighbor MT or LM warrant adding?} 
}

\section{Learned Keyless Joins for Context Enrichment}

Although the applications in Section~\ref{sec:applications} implicitly or explicitly perform the \emph{same} context enrichment, many state-of-the-art systems for these tasks re-implement and rediscover primitives across domains spanning machine learning, databases, information retrieval, and natural language processing~\citep{colbert,ditto,knnlm,rpt}.
We propose learned \textit{keyless joins} as an abstraction to unify context enrichment across these tasks and provide a vehicle to develop a no-code system with out-of-the-box recall exceeding na\"ive baselines.
In this section, we define keyless joins, provide a keyless join API specification, and introduce Transformer models that enable general keyless joins. 

\subsection{Keyless Joins: Definition and Objective}
\label{subsec:keyless}

Context enrichment requires a similarity-based join that operates at the record-, not schema-, level to retrieve related records without primary key-foreign key (KFK) relationships. 
We propose learned \emph{keyless joins} as an abstraction to enable this functionality.
The goal of a keyless join is to quantify the relatedness of records across different datasets to identify records referring to similar entities.

A keyless join must learn a common embedding space $\mathbf{X}$ for all records $d_{ij}$ across datasets $D_0 \cup D$ that reflects entity similarity.
For each $\mathbf{D_i}$, a keyless join learns a function $F_i: \mathbf{D_i} \rightarrow \mathbf{X}$ that maps elements of $D_i$ to the space $\mathbf{X}$.
We denote each transformed data point $F_i(d_{ij})$ as $x_{ij}$.
This mapping must be optimized such that related values map to similar feature vectors in $\mathbf{X}$ and unrelated values map to distant feature vectors in $\mathbf{X}$: $sim(x_{ij}, x_{kl}) > sim(x_{ij}, x_{mn})$ implies that the $j^{th}$ entity in the $i^{th}$ dataset is more closely related to the $l^{th}$ entity in the $k^{th}$ dataset than the $n^{th}$ entity in the $m^{th}$ dataset.
For instance, if we define similarity with respect to an $\ell_p$-norm, the above condition is equal to optimizing for  $\|x_{ij} -  x_{kl}\|_p < \|x_{ij} - x_{mn}\|_p$.


\begin{figure*}
\includegraphics[width=\linewidth]{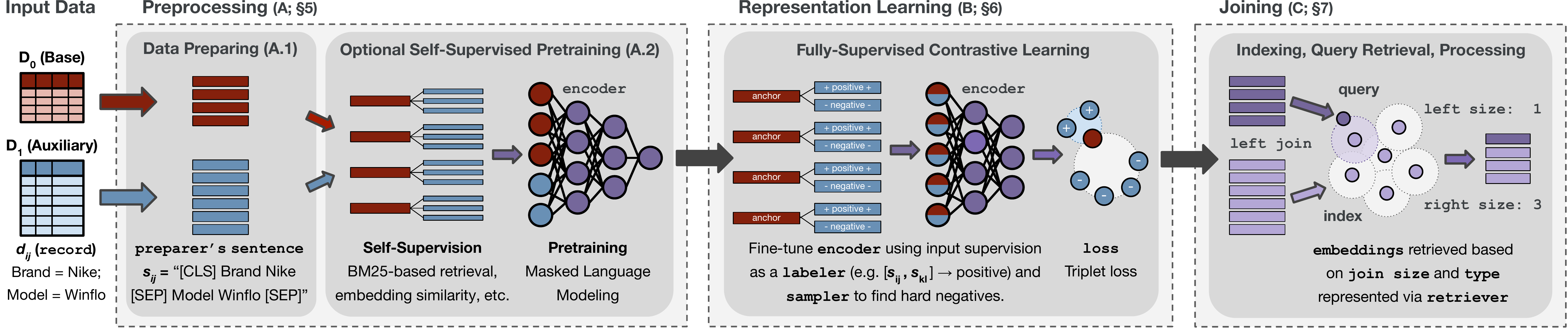}
\caption{\sys system architecture.}
\label{fig:pipeline}
\end{figure*}

\subsection{Join Specification for Context Enrichment}
\label{subsec:keyless_spec}

We propose a minimal join specification for our applications using keyless joins as a building block (Listing~\ref{listing:spec}).
Given a pair of data sources to join (\texttt{base\_table\_ref}, \texttt{aux\_table\_ref}), and examples of similar records across them (\texttt{supervision}), users first specify the \texttt{join type}: an inner join to only return enriched records, or outer join (left, right, full) to return enriched and unenriched records.
Users then specify the \texttt{join size}: how many records from one data source joins to a single record in the other data source to indicate one-to-one, one-to-many, or many-to-many semantics. 

As output, joined tuples (matches) between the two tables are the most similar records across them as learned using the keyless join objective. 
Results are ranked in order of greatest entity similarity, and the top k results are returned based on \texttt{join size}. 

\eat{
\begin{listing}[t]
\small
\begin{minted}{python}
base_table_ref [join_type] "KEYLESS JOIN" aux_table_ref 
"LEFT SIZE" integer "RIGHT SIZE" integer
"USING" supervision;

join_type = "INNER" | "LEFT" | "RIGHT" | "FULL";
\end{minted}
\caption{Keyless join specification and syntax, which defaults to an inner join.}
\label{listing:spec}
\end{listing}
}

\begin{lstlisting}[frame=single,caption={Keyless join specification (inner join default).},captionpos=b,float=t,label={listing:spec}, language=Python,commentstyle=\textcolor{teal},stringstyle=\dred,showstringspaces=false,belowskip=-2pt]
base_table_ref [join_type] "KEYLESS JOIN" aux_table_ref 
"LEFT SIZE" integer "RIGHT SIZE" integer
"USING" supervision;

join_type = "INNER" | "LEFT" | "RIGHT" | "FULL";
\end{lstlisting}

For instance, an entity matching application is written as follows to retrieve a single matching record between each data source:
\vsepfbox{%
    \parbox{.975\linewidth}{%
        \small{\texttt{%
            entity\_mentions\_A \dred{INNER KEYLESS JOIN} entity\_mentions\_B \\
            \dred{LEFT SIZE} 1 \dred{RIGHT SIZE} 1 \dred{USING} matching\_mentions;
        }}
    }%
}

\eat{\texttt{
\dred{entity\_mentions\_A} INNER KEYLESS JOIN \dred{entity\_mentions\_B}
LEFT SIZE \dred{1} RIGHT SIZE \dred{1}
}}

A search application is written as follows to retrieve 10 documents for each search query, else return the unenriched query: \\
\vsepfbox{%
    \parbox{.975\linewidth}{%
        \small{\texttt{%
        query\_corpus \dred{LEFT KEYLESS JOIN} document\_corpus \\
        \dred{LEFT SIZE} 1 \dred{RIGHT SIZE} 10 \dred{USING} relevant\_docs\_for\_query;
        }}
    }%
}

A recommendation application is written as follows to retrieve 10 items for each user, and 20 users who like each item: \\
\vsepfbox{%
    \parbox{.975\linewidth}{%
        \small{\texttt{%
        user\_database \dred{INNER KEYLESS JOIN} product\_database \\
        \dred{LEFT SIZE} 20 \dred{RIGHT SIZE} 10 \dred{USING} relevant\_docs\_for\_query;
        }}
    }%
}

In the remainder of paper, we describe our prototype system that implements this keyless join abstraction layer for enrichment.

\subsection{Background: Transformer-Based Encoders}
\label{subsec:transformer}

Our main tools for creating a representation optimized for the objective in Section~\ref{subsec:keyless} are Transformers~\citep{transformers}, as they have demonstrated success across a range of structured, semi-structured, and unstructured domains~\citep{tabert,bert,gpt2,neuraldb,picket}.
They consist of an encoder-decoder architecture, where the Transformer first encodes the input to an intermediate representation, and then decodes this representation to produce the end output.
Stacking encoder (e.g., BERT~\citep{bert}) or decoder (e.g., GPT-2~\citep{gpt2}) modules allows us to learn high-quality word embeddings that can either be used for a wide array of downstream applications. 
We focus on BERT-based embeddings. 

BERT embeddings used for a downstream task are trained in a two-step procedure. 
The first step is self-supervised pretraining using a masked language model (MLM) objective: a random subset of tokens are masked, and the model must predict each masked token given the surrounding text as context. 
Pretraining is performed with general purpose corpora such as Wikipedia. 
The second step is task-specific fine-tuning. 
Additional layers are appended to the final layer of the pretrained model based on the downstream task, and all model parameters are updated given small amounts of downstream supervision. 
Rather than relying solely on fine-tuning, an additional MLM pretraining step can be performed prior to fine-tuning to introduce additional domain knowledge. 

\begin{table}[]
\small
\begin{tabular}{lll}
\hline
\textbf{Data Types}                                                                                                                                              &                                                                                                                                                                                                                                                                                                                                                                                                               &                                                                                            \\ \hline
\textbf{\begin{tabular}[c]{@{}l@{}}\texttt{record}\\ \texttt{sentence}\\ \texttt{embedding}\end{tabular}}                                                             & \begin{tabular}[c]{@{}l@{}}\texttt{Dict[str, Union[str, int, float]]}\\ \texttt{str}\\ \texttt{List[float]}\end{tabular}                                                                                                                                                                                                                                                                                      &                                                                                            \\ \hline
\textbf{Operators}                                                                                                                                               &                                                                                                                                                                                                                                                                                                                                                                                                               & \multicolumn{1}{c}{\textbf{Step}}                                                          \\ \hline
\textbf{\begin{tabular}[c]{@{}l@{}}\texttt{preparer}\\ \texttt{labeler}\\ \texttt{sampler}\\ \texttt{loss}\\ \\ \texttt{encoder} \\ \texttt{retriever}\end{tabular}} & \begin{tabular}[c]{@{}l@{}}\texttt{record \textrightarrow{} sentence}\\ \texttt{Tuple[record, record] \textrightarrow{} bool}\\ \texttt{List[Generic[T]] \textrightarrow{} Generic[T]}\\ \texttt{Tuple[emb., Union[emb.,} \\ \hspace{1.75cm} Tuple[emb., emb.]]] \texttt{\textrightarrow{} float}\\ \texttt{sentence \textrightarrow{} embedding}\\ \texttt{List[emb.] \textrightarrow{} List[Tuple[record, emb.]]}\end{tabular} & \multicolumn{1}{c}{\begin{tabular}[c]{@{}c@{}}1\\ 1, 2\\ 2\\ 1, 2\\ \\ 1, 2 \\ 3\end{tabular}} \\ \hline
\end{tabular}
\caption{ \sys data types and operators used in each step}
\label{table:types}
\vspace{-1em}
\end{table}

\section{\sys}
\label{sec:ember}

We develop \sys,\footnote{https://github.com/sahaana/ember} an open-source system for no-code context enrichment.
\sys implements a keyless join abstraction layer that meets the specification in Section~\ref{subsec:keyless_spec}.
\sys first represents input records using transformer-based embeddings directly optimized for the condition in Section~\ref{subsec:keyless}.
\sys then populates a reusable index with these embeddings based on the \texttt{join type}, and configures index retrieval based on the \texttt{join sizes}. 
We now provide an overview of \sys's usage, API and architecture; data types and operators are in Table~\ref{table:types}, with an architecture overview in Figure~\ref{fig:pipeline}.

\subsection{Usage}
\label{subsec:api}

As input, \sys requires the join specification parameters: a base data source $D_0$ ("left"), auxiliary data sources $D=\{D_1,...,D_M\}$ ("right"), labeled examples of related data points that represent similar entities, \texttt{join type}, and \texttt{join sizes}.
Labeled examples are provided in one of two supervision forms: pairs of related records, one from each table, or a triple where a record from $D_0$ is linked to a related and unrelated record from each $D_i$.
That is, unrelated examples are optional, but related ones are required.
Recall that we focus on the $M = 1$ case, but describe extensions in Sections~\ref{sec:preprocessing}-\ref{sec:joining}.

As output, \sys retrieves the data points enriched based on the \texttt{join type} and \texttt{join sizes} as a list of tuples.
Users configure \sys using a json configuration file that exposes the join specification and lower-level \sys-specific parameters (see Section~\ref{subsec:extensibility}).


\subsection{API and Architecture}
\label{subsec:arch}

Keyless joins learn a representation that quantifies data point relatedness, and retrieve these points regardless of the input schema. 
We propose a modular system with three architectural elements to enable this: data preprocessing, representation learning, and data joining.
\sys consists of dataflow operators that transform inputs into the formats required for each step (see Table~\ref{table:types}). 

\sys represents input data records $d_{ij}$ as {\records} in key-value pairs.
Supervision is represented via \labelers.
An \sys pipeline consists of {\preparers}, {\encoders}, \samplers, \losses, and \retrievers. 
\sys uses \preparers to transform \records into \sentences.
\sys uses \encoders to transform \sentences into \embeddings that are optimized per a provided \loss; \samplers mine negative examples if the provided supervision only provides examples of related records.
The trained \embeddings are stored in an index, and are retrieved using a \retriever.
We provide an API over these operators in the form of a customizable, high-level configuration, as described in Section~\ref{subsec:easy}.
To enable functionality beyond that described, users must implement additional \preparers, \encoders, \samplers, or \retrievers. 
Users configure \sys using pre-defined or custom operators, or use the following pre-configured default:  

\minihead{Preprocessing (Figure~\ref{fig:pipeline}A, \S\ref{sec:preprocessing})} \sys ingests any text or numeric data source with a pre-defined schema, and converts its records to a common representation. 
By default, \sys converts \records to \texttt{sentence}s using \texttt{sentence preparer} modules, which are fed as input to the pipeline's \texttt{encoders}. 
\sys optionally pretrains the \texttt{encoders} in this step using self-supervision.

\minihead{Representation Learning (Figure~\ref{fig:pipeline}B, \S\ref{sec:representation})} \sys learns a mapping for each input data source's \records such that the transformed data is clustered by relatedness. 
To learn this mapping, \sys's \texttt{encoders} are fine-tuned with the input supervision (in the form of a \texttt{labeler}) and \texttt{loss}.
\sys applies the learned mapping to each of the \sentences to generate \embeddings passed to the final step.

\minihead{Joining (Figure~\ref{fig:pipeline}C \S\ref{sec:joining})} \sys populates an index with the learned \embeddings using Faiss~\citep{faiss}. 
A keyless join can be completed by issuing a similarity search query given an input \record (transformed to an \embedding) against this index.
The dataset that is indexed is determined by the \texttt{join type}. 
\sys uses a k-NN \texttt{retriever} module to retrieve as many records specified by the \texttt{join sizes}. 

\section{Preprocessing}
\label{sec:preprocessing}

In this section, we describe the first step of \sys's pipeline: preprocessing. 
Users provide base and auxiliary datasets with examples of related records across the two. 
\sys first processes the input datasets so all records are represented in a comparable format. 
\sys then pretrains the \encoders for the representation learning step. 
We describe these phases, and then multi-dataset extension.

\begin{figure}
\includegraphics[width=\columnwidth]{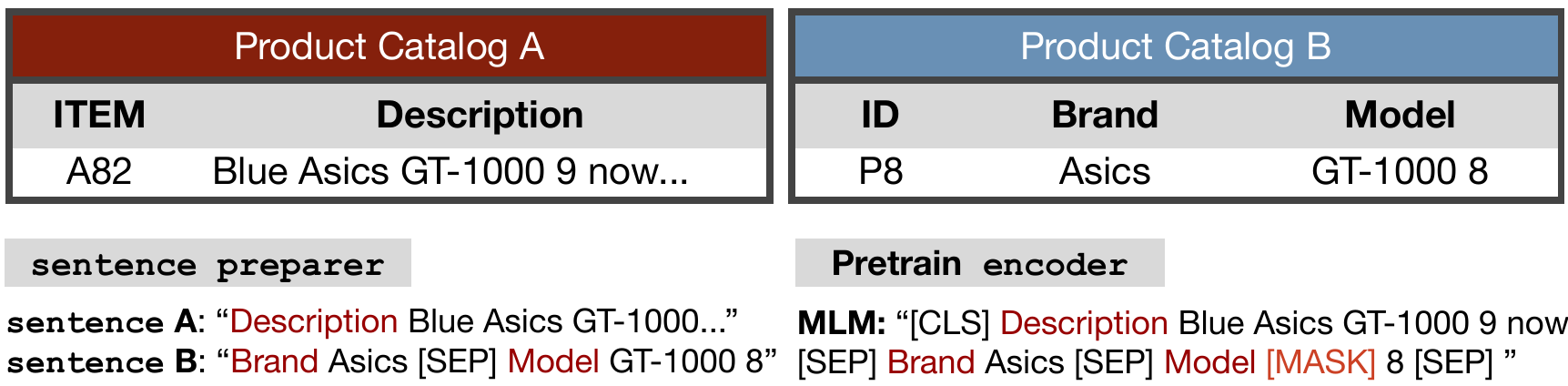}
\caption{Examples of the data preparing and pretraining phases of the preprocessing architecture step.}
\label{fig:preprocess}
\vspace{-1em}
\end{figure}

\minihead{Data Preparing (Figure~\ref{fig:pipeline}A.1)} This phase converts each \record $d_{ij}$ into a format that can be used for downstream Transformer-based \encoders regardless of schema.  
Motivated by recent work in representing structured data as sentences~\citep{tabert, neuraldb}, \sys's default pipeline uses a \sentence \preparer $P$ to convert input \records $d_{ij}$ into \sentences $P(d_{ij}) = s_{ij}$. 
Each key-value pair in $d_{ij}$ is converted to a string and separated by the \encoder's separator token as \texttt{``key\_1 value\_1 [SEP] key\_2 value\_2 [SEP]..."} (Figure~\ref{fig:preprocess}).

\minihead{Optional Pretraining (Figure~\ref{fig:pipeline}A.2)}
Our default BERT-based \encoder is trained over natural language corpora (Wikipedia and BookCorpus~\citep{wiki,bookcorpua}). 
Structured data is often domain-specific, and is of a different distribution than natural language.
Thus, we find that bootstrapping the pipeline's \encoders via additional pretraining can improve performance by up to \red{2.08$\times$}.
\sys provides a pretraining configuration option that is enabled by default.

Users can pretrain the pipeline's \encoders via BM25-based self-supervision (out-of-the-box) or by developing weak supervision-based \labelers (custom-built).
We add a standard Masked Language Modeling (MLM) head to each pipeline \encoder, and pretrain it following the original BERT pretraining procedure that fits a reconstruction loss. 
To  encourage the spread of contextual information across the two tables, we concatenate one \sentence from each table to one other as pretraining input as \texttt{``$s_{ij}$ [SEP] $s_{kl}$"} (Figure~\ref{fig:preprocess}). 
We select \sentence pairs that are likely to share information in an unsupervised manner via BM25, a bag-of-words relevance ranking function~\citep{bm25}, though any domain-specific unsupervised similarity join can be used, such as from Magellan~\citep{magellan} or AutoFJ~\citep{autofj}. 
We evaluate BM25-based MLM pretraining in Section~\ref{subsec:micros}.
During development, we explored learning conditional representations of each table given the other by selectively masking one table's \sentences, but found no change in performance compared to uniform masking. 

\minihead{Multi-Dataset Extension} 
Depending on the encoder configuration used in the multi-dataset case, the BM25-based pretraining step will be applied to all \sentences across all datasets at once (anchored from the base table), or each dataset pairwise. 

\section{Representation Learning}
\label{sec:representation}

In this section, we describe the second step of \sys's pipeline: representation learning.
Given the \sentences and optionally pretrained \encoders from the first step, \sys fine-tunes the \encoders such that \embeddings from related records are close in embedding space. 
These \embeddings are then passed to the final pipeline step. 
Users can choose how many \encoders to configure, the \encoder architecture and output dimension, and how to train the \encoders, which we now detail prior to describing multi-dataset extension.

\minihead{\texttt{Encoder} configuration} \sys provides a choice of training independent \encoders, $E_i$ for each data source, or using a single \encoder $E_0$ common to all data sources. 
In all of our scenarios, using a single, common \encoder performs best, sometimes by over an order of magnitude, and is set as \sys's default configuration.
In our tasks, we found that using separate encoders perturbs the representations such that exact matches no longer share a representation---thus, the encoders must relearn correspondences, and often fail to.
However, to extend \sys to non-textual data (e.g., images, video, text-image for image captioning or joint sentiment analysis, or text-audio tasks), users must leverage source-specific \encoders---thus, we provide them as a configurable option.

\minihead{\texttt{Encoder} architecture} \sys lets users configure each \encoder's base architecture and the size of the learned \embeddings. 
Users can choose BERT$_{\text{base}}$ or DistilBERT$_{\text{base}}$ as the core architecture; \sys's default is the 40\% smaller, and 60\% faster DistilBERT model.
We use models from HuggingFace Transformers~\citep{hf}; thus, integrating new architectures requires few additional lines of code.

We remove the MLM head used in the preprocessing step for optional pretraining, and replace it with a fully connected layer that transforms BERT's default 768-dimensional output to a user-specified \embedding output dimension, $d$. 
The output of the fully connected layer is a $d$-dimensional embedding for each input token. 
Users can choose one of two types of aggregation methods that will return a single $d$-dimensional output per input \sentence: averaging all of the embeddings, or using the embedding of the leading \texttt{CLS} token that BERT appends to its input data (Figure~\ref{fig:preprocess}). 
\sys defaults to \texttt{CLS}-based aggregation with a 200-dimensional output.

\minihead{\texttt{Encoder} training} The \encoders' goal is to learn a representation (\embedding), for the data sources such that \sentences that refer to similar entities are grouped in the underlying embedding space. 
Recall that to perform a keyless join under an $\ell_p$-norm, each \encoder must learn a function that maps elements of $D_i$ to the space $\mathbf{X}$, such that $\|x_{ij} -  x_{kl}\|_p < \|x_{ij} - x_{mn}\|_p$ when the $j^{th}$ entity in the $i^{th}$ dataset is more closely related to the $l^{th}$ entity in the $k^{th}$ dataset than the $n^{th}$ entity in the $m^{th}$ dataset.
We directly optimize for this objective function by training \encoders using a contrastive, triplet loss together with user-provided supervision.

Given an anchor \record  $d_{0\text{a}}$ from $D_0$, and  \records $d_{1\text{p}}$ and $d_{1\text{n}}$ in $D_1$ that are related and unrelated to $d_{0\text{a}}$, respectively, let $F_0(d_{0\text{a}}) = x_{0\text{a}}$, $F_1(d_{1\text{p}}) = x_{1\text{p}}$, and $F_1(d_{1\text{n}}) = x_{1\text{n}}$ be their \embeddings following the composition of the \sentence \preparer $P$ and \encoder $E_i$ operators ($F_i=E_i \circ P$).
We minimize the triplet loss, defined as:
$$ \mathcal{L}(x_{0\text{a}}, x_{1\text{p}}, x_{1\text{n}}) = \max\{\|x_{0\text{a}} - x_{1\text{p}}\|_p,  \|x_{0\text{a}} - x_{1\text{n}}\|_p + \alpha, 0\}$$

\noindent where $\alpha$ is a hyperparameter controlling the margin between related and unrelated \embeddings.
Users could use an alternative \loss function: a cosine similarity loss would provide representations that similarly encourage related points to be close and unrelated points to be far, and a standard binary cross-entropy loss may be used as well, where the final layer in the trained \encoder would be the end embeddings.
However, they do not explicitly optimize for \textit{relative} distances between related and unrelated records. 

Users often have examples of related pairs, but no unrelated examples to form triples.
In our tasks, only the search workload provides triples, while the rest provide lists of related pairs, which we represent with a \labeler. 
As a result, we provide negative \sampler operators to convert \labelers that operate over pairs to the triples required by the triplet \loss. 
If the user-provided \labeler already contains a related and unrelated example for a \record, we form a triple using these examples.
If a \record does not contain a related and unrelated example, for each \record $d_{0j}$ in $D_0$ with a labeled related record, we use either a random \sampler or a stratified hard negative \sampler. 
The random \sampler selects a \record at random (that is not a supervised related pair) as an example unrelated \record. The stratified \sampler provides tiers for different degrees of relatedness for the negative examples---a user can specify a region of hard examples to sample from, given prior domain knowledge.
We provide a default stratified \sampler with a single tier of relatedness defined using BM25 or Jaccard similarity, though any unsupervised join method can be used (including AutoFJ~\citep{autofj}). We show that a hard negative \sampler improves recall by up to \red{30\%} compared to a random \sampler in Section~\ref{subsec:micros}.

\minihead{Multi-Dataset Extension} The \encoder configuration and architecture are unchanged in the multi-dataset case. 
There are two options for the \encoder training procedure.
The first scenario follows the context enrichment problem definition---where a base table $D_0$ must be augmented with several auxiliary data sources $D$. 
In this case, each \encoder can be trained pairwise with $D_0$ and $D_i$ as described above.
In the case where multiple data sources must be aligned in sequence, a user can successively apply \sys over a data DAG created using their prior knowledge---effectively creating several, sequential context enrichment problems as described in Section~\ref{subsec:target} for multi-hop question answering. 
Else, \sys can induce a DAG and sequentially traverse the space of possible keyless joins based on the cardinality of each dataset, in ascending order.

\section{Joining}
\label{sec:joining}

In this section, we describe the last step of \sys's pipeline: joining. 
Given the \embeddings output by the trained \encoders, \sys executes the  keyless join by identifying related points across the input datasets, and processing them for downstream use.
\sys indexes the learned \embeddings and queries this index to find candidate related records.
The \texttt{join sizes} determine the number of returned results per query, and we also allow users to configure a similarity threshold for each candidate match.
We now detail the indexing and retrieval procedure, post-processing, and multi-dataset extension.

\minihead{Indexing and Query Retrieval}
Given two records, \sys computes the similarity between their \embeddings to determine if the records are related. 
Many traditional solutions to our motivating applications perform such pairwise checks across all possible pairs either na\"ively or with blocking~\citep{deepmatcher, blocking, bertmarco} to identify related records. 
However, the choice of operator is both domain-specific, and scales quadratically at query time with the size of the input datasets and blocking mechanisms used (see Section~\ref{subsec:general}).
We eliminate the need for pairwise checking by indexing our \embeddings, which are optimized for clustering related records, and rely on efficient libraries for maximum inner product search (MIPS)~\citep{faiss}.

For a \texttt{LEFT} or \texttt{RIGHT OUTER JOIN}, \sys constructs an index over the base ($D_0$) or auxiliary ($D_1$) datasets, respectively. 
\sys then queries the index with each \embedding of the remaining dataset, and returns the \record and \embedding corresponding to the most similar records using a \retriever operator.
For a \texttt{FULL OUTER} or \texttt{INNER JOIN}, \sys may jointly index and query both datasets to identify related entries in either direction. 
By default, \sys only indexes the larger dataset to reduce runtime---an optimization we evaluate in Section~\ref{subsec:micros} that improves query runtime by up to \red{2.81}$\times$.
\sys's default configuration is an \texttt{INNER JOIN}.


\minihead{Post-Processing}
The user-provided \texttt{join size} configures the \retriever to return the top-$k$ records with the closest embeddings in the indexed dataset.
\sys additionally supports threshold-based retrieval.
The former is useful for applications such as search, where a downstream task may be to display the top-$k$ search results to a user in sorted order.
The latter is useful for applications where there may not be related records for each record in the base dataset. 
\sys's default \retriever is configured for a 1-to-10 join.

If the task $T$ is not to simply return related records, users can construct pipelines on top of \sys's outputs, relying on keyless-join-based context enrichment as a key primitive.
Examples of such workloads include the recommendation example from Figure~\ref{fig:example}, which we simulate as a case study in Section~\ref{subsec:extensibility}, open domain question answering, data augmentation, or applications that involve humans in the loop to verify matches or drive business needs. 

\minihead{Multi-Dataset Extension} The joining step may vary in the multi-dataset case in two ways. 
In a first scenario, each auxiliary data source $D_i$ is indexed, and the query retrieval phase will query each of these indexes and return candidate related data points for each pairwise interaction. 
If multiple data sources must be sequentially aligned and a DAG can be specified over this sequence, a user can chain context enrichment subroutines by querying the next index with the records retrieved by the previous subroutine; we provide a toy example of this scenario in Section~\ref{subsec:extensibility} under Recommendation. 

\section{Evaluation}
\label{sec:eval}

In this section, we demonstrate that \sys and its operators are:

\begin{enumerate}
    \item \textbf{General}: \sys enables context enrichment across five domains while meeting or exceeding similarity-join baseline recall and query runtime performance (Section~\ref{subsec:general}). 
    \item \textbf{Extensible}: \sys provides a modular architecture, where each component affects performance (Section~\ref{subsec:micros}). \sys enables task-specific pipelines for various similarity-based queries, and provides task performance that can be fine-tuned by state-of-the-art systems (Section~\ref{subsec:extensibility}). 
    \item \textbf{Low Effort}: \sys requires no more than five configuration changes (Table~\ref{table:datasets}) from its default, and does not always require large amounts of hand-labeled examples (Section~\ref{subsec:easy}).
\end{enumerate}

\begin{table}[]
\resizebox{\columnwidth}{!}{%
\begin{tabular}{|c|lrrrrr|}
\hline
\textbf{\textbf{Task}} & \textbf{\textbf{Dataset}} & \textbf{\textbf{\# base}} & \textbf{\textbf{\# aux}} & \textbf{\textbf{\# + train}} & \textbf{\textbf{\# + test }} & \textbf{\textbf{\# LoC}} \\ \hline
FJ                     & IMDb                      & 50000                     & 10000                     & 40000                     & 10000                     & 1                    \\ \hline
FJ                     & IMDb-hard                 & 50000                     & 10000                     & 40000                     & 10000                     & 1                    \\ \hline
QA                     & SQuAD                     & 92695                     & 64549                    & 86668                      & 6472                      & 2                        \\ \hline
S                      & MS MARCO                  & 508213                    & 8.8M                     & 418010                     & 7437                      & 1                       \\ \hline
R                      & IMDb-wiki                 & 47813                     & 47813                    & 38250                     & 9563                      & 1 (6)                       \\ \hline
EM-T                   & Abt-Buy                   & 1081                      & 1092                     & 616                        & 206                       & 2                      \\ \hline
EM-T                   & Company                   & 28200                     & 28200                    & 16859                      & 5640                      & 2                      \\ \hline
EM-S                   & BeerAdvo-RateBeer         & 4345                      & 3000                     & 40                         & 14                        & 1                        \\ \hline
EM-S                   & iTunes-Amazon             & 6907                      & 55923                    & 78                         & 27                        & 1                        \\ \hline
EM-S                   & Fodors-Zagat              & 533                       & 331                      & 66                         & 22                        & 1                        \\ \hline
EM-S                   & DBLP-ACM                  & 2616                      & 2294                     & 1332                       & 444                       & 1                        \\ \hline
EM-S                   & Amazon-Google             & 1363                      & 3226                     & 699                        & 234                       & 1                        \\ \hline
EM-S                   & DBLP-Scholar              & 2616                      & 64263                    & 3207                       & 1070                      & 1                        \\ \hline
EM-S                   & Walmart-Amazon            & 2554                      & 22074                    & 576                        & 193                       & 1                        \\ \hline
EM-D                   & DBLP-ACM                  & 2616                      & 2294                     & 1332                       & 444                       & 1                        \\ \hline
EM-D                   & DBLP-Scholar              & 2616                      & 64263                    & 3207                       & 1070                      & 1                        \\ \hline
EM-D                   & iTunes-Amazon             & 6907                      & 55923                    & 78                         & 27                        & 1                        \\ \hline
EM-D                   & Walmart-Amazon            & 2554                      & 22074                    & 576                        & 193                       & 1                        \\ \hline
\end{tabular}%
}
\caption{Data source record count, related pair (supervision) count, and number of configuration lines changed for the best recall@k, and for downstream tasks in parenthesis.}
\label{table:datasets}
\vspace{-10pt}
\end{table}

\subsection{Evaluation Metric and Applications}

\minihead{Context Enrichment Evaluation Metric} Context enrichment identifies related records across a base and auxiliary dataset. 
We can view related records between the two datasets as forming edges in a bipartite graph: each record in each dataset represents a vertex in the graph. 
Under this framing, a context enrichment system must retrieve all of the outgoing edges (i.e., related records) for each record in the base dataset.
This is equivalent to maximizing record-level recall, or the fraction of records for which we recover \emph{all} related records. 
We choose to define recall at the record-level, rather than the edge-level, as we view context enrichment systems as being repeatedly queried for new incoming data once instantiated. 

A na\"ive means to optimize for recall is to return \textit{all} auxiliary records as being related to each record in the base dataset. 
A precision metric greatly drops when we retrieve multiple records.
Thus, our evaluation metric is \textbf{recall@k}, for small k (i.e., \texttt{join size}).

\minihead{Applications} We evaluate \sys against workloads from five application domains: fuzzy joining, entity matching, search, question answering, and recommendation (summarized in in Table~\ref{table:datasets}). 
We make all our datasets publicly available post-\sys processing. 

\subsubsection{Fuzzy Join (FJ)}
We build two workloads using a dataset and generation procedure from a 2019 scalable fuzzy join VLDB paper~\citep{surajit}. 
The first dataset consists of the Title, Year, and Genre columns from IMDb~\citep{imdb}.
The second dataset is generated by perturbing each row in the first by applying a combination of token insertion, token deletion, and token replacement. 
The task $T$, is to join each perturbed row with the row that generated it.
We generate two dataset versions: 5 perturbations per row (IMDb) and 15 perturbations per row (IMDb-hard), up to 25\% of the record length. 
As we focus on generalizablity more than scalability, to form a dataset, we randomly sample 10,000 movies and generate 5 perturbed rows for each. 
We hold out 20\% of records as the test set; no records from the same unperturbed record are in both the train and test sets.

\subsubsection{Entity Matching (EM)}
We use all 13 benchmark datasets~\citep{dmdata} released with DeepMatcher~\citep{deepmatcher}, spanning structured (EM-S), textual (EM-T), and dirty (EM-D) entity matching. 
The base and auxiliary datasets always share the same schema. 
In EM-T, all data records are raw text entries or descriptions. 
In EM-S, each data record is drawn from a table following a pre-defined schema, where text-based column values are restricted in length. 
In EM-D, records are similar to EM-S, but some column values are injected into the incorrect column.
The task $T$ is to label a record pair (one from each dataset) as representing the same entity or not. 
Train, validation, and test supervision are lists of unrelated, and related pairs---we only use the related pairs for training \sys, as we identified mislabeled entries (false negatives) when using \sys to explore results.

\subsubsection{Search (S)}
We use the MS MARCO passage retrieval benchmark~\citep{marco}.
MS MARCO consists of a collection of passages from web pages that were gathered by sampling and anonymizing Bing logs from real queries~\citep{marco}. 
The task $T$ is to rank the passage(s) that are relevant to a given query as highly as possible. 
Supervision is a set of 397M triples, with 1 relevant and up to 999 irrelevant passages for most queries.
Irrelevant passages are retrieved via BM25.
We report results over the publicly available labeled development set. 

\subsubsection{Question Answering (QA)}
We modify the Stanford Question Answering Dataset (SQuAD)~\citep{squad}.
SQuAD consists of Wikipedia passages and questions corresponding to each passage. 
The task $T$ is to identify the beginning of the text span containing the answer to each question. 
As described in Section~\ref{sec:applications}, a retriever module, used in retriever-reader models for QA~\citep{retriever}, performs context enrichment.
We modify SQuAD by splitting each each passage at the sentence-level, and combining these sentences to form a new dataset. 
The modified task is to recover the sentence containing the answer.

\subsubsection{Recommendation (R)}
We construct a workload using IMDb and Wikipedia to mimic the e-commerce example from Figure~\ref{fig:example}.
For the first dataset, we denormalize four IMDb tables using KFK joins: movie information (title.basics), principal cast/crew for each movie (title.principals) and their information (name.basics), and movie ratings (title.ratings)~\citep{imdb}.
For the second dataset, we extract the summary paragraphs of the Wikipedia entry for each IMDb movie by querying latest Wikipedia snapshot; we extract 47813 overlapping records~\citep{wiki}.
We remove the IMDb ID that provides a KFK relationship to induce a need for keyless joins.
We construct this workload to enable two applications. 
In Application A, we show that \sys can join datasets with dramatically different schema. 
In Application B, we show how to estimate the rating for movies in the test set given ratings for the train set by performing similarity-based analyses enabled by \sys. 
Supervision is provided as exact matches and movie ratings with an 80-20 train-test set split.

\begin{table*}[]
\resizebox{\textwidth}{!}{%
\begin{tabular}{ll|c|c|c|c|c|c|c|c||c|c|c|c|c|c|c|c|}
\hhline{~~----------------}
                                                   &                                 & \multicolumn{8}{c||}{\cellcolor[HTML]{EFEFEF}\textbf{Recall@1}}                                                                                                                                                                                                                                          & \multicolumn{8}{c|}{\cellcolor[HTML]{EFEFEF}\textbf{Recall@10}}                                                                                                                                                                                                                         \\ \hline
\rowcolor[HTML]{F3F3F3} 
\multicolumn{1}{|l|}{\cellcolor[HTML]{EFEFEF}Task} & \cellcolor[HTML]{EFEFEF}Dataset & \textbf{\textsc{LD}}                   & \textbf{\textsc{J-2g}}                  & \textbf{\textsc{JK-2g}}                & \textbf{\textsc{JK-ws}} & \textbf{\textsc{J-ws}} & \textbf{\textsc{BM25}}                 & \textbf{\textsc{bert}} & \cellcolor[HTML]{A4C8E9}\textbf{\textsc{emb}} & \textbf{\textsc{LD}}                   & \textbf{\textsc{J-2g}}                  & \textbf{\textsc{JK-2g}}                & \textbf{\textsc{JK-ws}} & \textbf{\textsc{J-ws}} & \textbf{\textsc{BM25}} & \textbf{\textsc{bert}} & \cellcolor[HTML]{A4C8E9}\textbf{\textsc{emb}} \\ \hline
\multicolumn{1}{|l|}{FJ}                           & IMDb                       & \cellcolor[HTML]{DCEAF7}\textbf{99.32} & \cellcolor[HTML]{DCEAF7}\textbf{100.00} & \cellcolor[HTML]{DCEAF7}\textbf{99.42} & 71.54                   & 91.61                  & 91.81                                  & 45.85                  & 97.86                                         & \cellcolor[HTML]{DCEAF7}\textbf{99.99} & \cellcolor[HTML]{DCEAF7}\textbf{100.00} & \cellcolor[HTML]{DCEAF7}\textbf{99.96} & 96.13                   & 96.30                  & 96.33                  & 69.83                  & \cellcolor[HTML]{DCEAF7}\textbf{99.79}        \\ \hline
\multicolumn{1}{|l|}{FJ}                           & IMDb-hard                       & 95.54                                  & \cellcolor[HTML]{DCEAF7}\textbf{98.87}  & 95.02                                  & 33.42                   & 52.29                  & 54.58                                  & 8.72                   & 88.92                                         & \cellcolor[HTML]{DCEAF7}\textbf{99.04} & \cellcolor[HTML]{DCEAF7}\textbf{99.82}  & 98.26                                  & 42.68                   & 64.34                  & 65.82                  & 21.11                  & 98.37                                         \\ \hline
\multicolumn{1}{|l|}{QA}                           & SQuAD                           & 1.37                                   & 34.70                                   & 41.26                                  & 29.47                   & 27.26                  & 49.17                                  & 11.37                  & \cellcolor[HTML]{DCEAF7}\textbf{52.91}        & 1.50                                   & 52.82                                   & 61.18                                  & 44.79                   & 43.18                  & 67.08                  & 27.03                  & \cellcolor[HTML]{DCEAF7}\textbf{78.85}        \\ \hline
\multicolumn{1}{|l|}{S}                            & MS MARCO                        & 0.26                                   & 1.66                                    & 1.66                                   & 1.15                    & 1.15                   & 2.31                                   & 0.01                   & \cellcolor[HTML]{DCEAF7}\textbf{16.34}        & 0.96                                   & 7.38                                    & 7.38                                   & 5.11                    & 5.11                   & 4.10                   & 0.10                   & \cellcolor[HTML]{DCEAF7}\textbf{46.98}        \\ \hline
\multicolumn{1}{|l|}{R}                          & IMDb-wiki                       & 58.96                                  & 11.01                                   & 82.57                                  & 86.08                   & 18.30                  & 63.25                                  & 0.04                   & \cellcolor[HTML]{DCEAF7}\textbf{97.02}        & 64.62                                  & 18.80                                   & 93.12                                  & 91.93                   & 18.30                  & 96.25                  & 0.26                   & \cellcolor[HTML]{DCEAF7}\textbf{98.89}        \\ \hline
\multicolumn{1}{|l|}{EM-T}                         & Average                         & 18.20                                  & 20.06                                   & 36.56                                  & 36.90                   & 37.21                  & 61.09                                  & 6.88                   & \cellcolor[HTML]{DCEAF7}\textbf{71.93}        & 25.73                                  & 37.57                                   & 49.00                                  & 56.15                   & 53.11                  & 71.90                  & 21.45                  & \cellcolor[HTML]{DCEAF7}\textbf{83.29}        \\ \hline
\multicolumn{1}{|l|}{EM-S}                         & Average                         & 52.26                                  & 73.50                                   & 72.30                                  & 69.25                   & 72.71                  & \cellcolor[HTML]{DCEAF7}\textbf{76.59} & 38.98                  & \cellcolor[HTML]{DCEAF7}\textbf{77.20}        & 70.26                                  & 94.41                                   & \cellcolor[HTML]{DCEAF7}\textbf{97.90} & 93.51                   & 93.33                  & 95.19                  & 55.88                  & \cellcolor[HTML]{DCEAF7}\textbf{97.58}        \\ \hline
\multicolumn{1}{|l|}{EM-D}                         & Average                         & 18.70                                  & 64.69                                   & 52.05                                  & 53.41                   & 61.94                  & 65.47                                  & 30.61                  & \cellcolor[HTML]{DCEAF7}\textbf{66.56}        & 24.60                                  & 90.86                                   & 80.92                                  & 79.41                   & 89.85                  & 92.31                  & 43.92                  & \cellcolor[HTML]{DCEAF7}\textbf{97.83}        \\ \hline
\end{tabular}%
}
\caption{Baseline comparison, where we highlight all methods within 1\% of the best for each task. No single baseline dominates the other, but Ember (\textsc{emb}) is competitive with or better than alternatives in nearly all tasks with respect to Recall@k.}
\label{tab:baselines}
\vspace{-15pt}
\end{table*}


\eat{\minihead{Evaluation Metrics} We evaluate all EM workloads with respect to F1 score, as is standard in existing work~\citep{deepmatcher,ditto}.
For the remainder of the workloads, we report MRR@k and Retrieval@k for k = 1, 10. 
MRR@k refers to the average of the multiplicative inverse of the rank of the first relevant item in the top k ranked items, and is an information retrieval standard~\citep{marco}.
We define Retrieval@k to be the percentage of the test set whose true match is present in the top-k returned nearest neighbors by the \retriever. 
We report Retrieval@k as a modification of Recall@k, to reflect that we are only looking for a single match in these join scenarios.
MRR@1 is equivalent to Retrieval@1.}

\subsection{Experimental Setup}
\label{sec:setup}

\minihead{Baselines} We first evaluate all workloads compared to seven similarity-based joins with respect to recall@1 and recall@10. 
Our baselines are joins using Levenshtein distance, four variations of Jaccard Similarity, BM25, Auto-FuzzyJoin~\citep{autofj}, and a pretrained embedding-based approach. 
We evaluate downstream EM and search workloads with respect to previously reported state-of-the-art and benchmark solutions~\citep{deepmatcher,deeper,ditto,marcohomepage,bertmarco}.
The remainder of our workloads were constructed to isolate context enrichment from the downstream task, and do not have standard baselines. 

\minihead{\sys Default Configuration} Tasks use a \sentence \preparer, and perform 20 epochs of self-supervised pretraining.
\sys uses a single DistilBERT$_{\text{base}}$ \encoder trained with a triplet \loss and \red{stratified} hard negative \sampler. 
Self-supervision and the \sampler use BM25 to identify similar \sentences to concatenate from each data source, and to mark as hard negatives, respectively. 
The output \embedding size is 200, and models are trained with a batch size of 8 using an ADAM optimizer~\citep{adam} with initial learning rate of $1\mathrm{e}{-5}$.
The default \texttt{join sizes} are 1 and 10.
Results are five-trial averages.

\minihead{Implementation} We use a server with two Intel Xeon Gold 6132 CPUs (56 threads) with 504GB of RAM, and four Titan V GPUs with 12GB of memory.
We implement \sys in Python and PyTorch~\citep{pytorch}, with pretrained models and tokenizers from HuggingFace Transformers~\citep{hf}. 
We use Faiss for MIPS~\citep{faiss}, Magellan for similarity join baselines~\citep{magellan}, rank-BM25 for BM25~\citep{bm25_code}, and AutoFuzzyJoin~\citep{autofj}.

\subsection{Generalizability}
\label{subsec:general}

We show that \sys's recall and runtime meets or outperforms that of the following similarity-join baselines in nearly all tasks:

\minihead{Levenshtein-Distance (LD)} LD is the number of character edits needed to convert one string to another. 
This join returns the closest records with respect to single-character edits over provided key columns. We filter and only return results under a 30 edit threshold.

\minihead{Jaccard-Similarity, Specified Key (JK-WS, JK-2G)} The Jaccard similarity between sets $A$ and $B$ is $J(A, B) = \frac{|A\cap B|}{|A \cup B|}$.
Defining a Jaccard-similarity based join over textual inputs requires tokenizer selection.
We consider a whitespace tokenizer (WS) and a 2-gram tokenizer (2G) to capture different granularities. 
JK-WS and JK-2G return the closest records with respect to Jaccard similarity over provided key columns using a WS or 2G tokenizer.
We set a filtering threshold to return results with at least 0.3 Jaccard similarity.

\minihead{Jaccard-Similarity, Unspecified Key (J-WS, J-2G)} J-WS and J-2G return the closest records with respect to Jaccard similarity using a WS or 2G tokenizer, after a \sentence \preparer.
We set a filtering threshold to return results with over 0.3 Jaccard similarity.

\minihead{BM25 (BM25)} BM25 is a bag-of-words ranking function used in retrieval~\citep{bm25}. This join returns the closest records with respect to the Okapi BM25 score using default parameters k1=1.5 and b=0.75.

\minihead{Pretrained-Embedding (BERT)}
BERT generates embeddings for each prepared \sentence via a pretrained DistilBERT$_{\text{base}}$ model, and returns the closest records based on the $\ell_2$-norm between them. 

\red{\minihead{Auto-FuzzyJoin (AutoFJ)}
AutoFJ automatically identifies join functions for unsupervised similarity joins~\citep{autofj} by assuming one input is a "reference table" with few or no duplicates, which does not always hold in context enrichment.
If duplicates are present, or records in the reference table are not sufficiently spread, precision estimation may break down (\sys trivially accounts for these scenarios).
AutoFJ optimizes for a precision target, and does not expose any \texttt{join size} semantics.
We find a single result is typically returned per record, so we only consider recall@1 at a low precision target of 0.5. 
We comment on workloads that completed in 1.5 days on our machines using all cores. 
As AutoFJ exhaustively computes record similarities for each considered join configuration, this means we omit results for MS MARCO, EM-T Company, EM-S BeerAdvo-RateBeer, and several others consisting of large text spans, or where blocking may not be as effective.
Users must manually align \emph{matching} columns, which do not always exist thus ignore information in our tasks (e.g., IMDb-wiki): as input, we pass the output of a \sentence \preparer if column names are not identical.
}
\\

\noindent For multi-column datasets, we provide a plausible key column if the method requires.
In FJ, EM, and R, this is a title or name column. 

\subsubsection{Retrieval Quality} We show that \sys is competitive with or exceeds the best performing alternatives with respect to Recall@1 and Recall@10 (Table~\ref{tab:baselines}). 
In both tables, we highlight the methods that are within 1\% of the top performing method. 
No method dominates the others across the different workloads, while BERT underperformed all alternatives.  
Broadly, character-based joins (LD, JK-2G, J-2G) tend to perform well in scenarios where a join key exists but may be perturbed (FJ, EM-S, EM-D); word-based joins (J-WS, BM25) tend to perform well in scenarios with word-level join keys (EM-S), or where a join key does not exist but common phrases still link the two datasets (R, EM-D).
Neither performs well otherwise (S, EM-T).
\sys most consistently outperforms or is comparable with alternatives across the tasks: a learned approach is necessary for generalization. 
The exception is with the synthetic FJ workload that is, by construction, suited for character-based joins. 

\red{We do not report AutoFJ in Table~\ref{tab:baselines} due to incompleteness. 
AutoFJ is tailored for the EM and FJ: many-to-one joins with identical schema. 
Of those completed, \sys is sometimes outperformed by up to 8\% recall@1 (Amazon-Google), but is otherwise comparable or far exceeds AutoFJ when considering recall@10.
AutoFJ's recall is 37.9\% for QA, and 38.5\% for R.
As AutoFJ over the Wikipedia summary (task R) did not terminate, we only use the title column.
}

\subsubsection{Query Runtime} Using optimized MIPS routines improved query runtime performance by up to two orders of magnitude.
We note that this reflects \sys joining time, not pretraining and representation learning, which may require several hours.
As the two FJ datasets are synthetic and follow identical generation methods, we consider just IMDb-hard.
For our largest dataset, MS MARCO (S) with 8.8M auxiliary records, \sys indexes and retrieves results from the entire corpus, but each baseline only considers each of the up to 1000 relevant records provided by the MS MARCO dataset for each base record. 
Even when using CPU-only Faiss~\citep{faiss}, \sys takes \red{7.24} seconds on average for the MIPS-based joining step. 
Embedding the query table takes \red{24.52} seconds on average.
The most expensive dataset with respect to MIPS runtime is MS MARCO (S), which required \red{1.97} minutes when indexing all 8.8M records; excluding it reduces the average to \red{0.31} seconds.
For embedding time, this was EM-T Company, requiring \red{3.96} minutes, whose exclusion reduces the average to \red{11.19} seconds.
In contrast, the fastest baseline method (JK-WS), took \red{5.02} minutes on average; excluding MS MARCO (\red{79.73} minutes) reduces the average to \red{21.18} seconds.
The slowest method (LD), took \red{23.64} minutes on average, with the most expensive dataset, EM-T Company, requiring \red{321.98} minutes.

\subsection{Extensibility: Architecture Lesion}
\label{subsec:micros}
We demonstrate modularity and justify our architectural defaults by evaluating \sys's performance when modifying components as follows, each of which requires a single line configuration change (all reported percentages are relative unless explicitly specified): 
\begin{enumerate}
    \item Replacing MLM pretraining and representation learning (RL) with a pretrained \encoder.
    \item Removing RL, leaving only MLM pretraining.
    \item Replacing the fine-tuned Transformer model with a frozen pretrained embedding and a learned dense layer.
    \item Removing MLM pretraining.
    \item Random negatives compared to hard negative sampling
    \item Using an \encoder for each dataset versus a single \encoder.
    \item Removing the joining index-query optimization.
\end{enumerate}

\subsubsection{Pretrained \encoder (\textsc{-mlm, rl})}
We remove MLM pretraining and RL, and use a pretrained DistilBERT$_{\text{base}}$ encoder---i.e., the same as BERT in Table~\ref{tab:baselines}. 
We report our results in Figure~\ref{fig:lesion} normalized by \sys's default. 
\sys's performance dramatically declines, sometimes by three orders of magnitude (e.g., recall dropped to 0.04 in R).
It is only feasible when the contents of the datasets to be joined are similar, as in certain EM-S and EM-D tasks, and FJ-IMDb.

\subsubsection{Removing Representation Learning (\textsc{-rl})}
We remove representation learning (RL) and only pretrain an \encoder using the BM25-based MLM procedure from Section~\ref{sec:preprocessing}.
We report our results in Figure~\ref{fig:lesion}. 
\sys's performance declines by up to an order of magnitude, though it improves performance compared to \textsc{-mlm, rl} on tasks whose datasets do not match the original natural text Wikipedia corpus (e.g., EM-S, FJ). 
Primarily textual datasets do not see as large an improvement over \textsc{-mlm, rl}, and for QA, where the corpus is derived from Wikipedia, \textsc{-rl} performs slightly worse.

\subsubsection{Remove Transformer fine-tuning (\textsc{-ft})}
We replace RL with fixed MLM pretrained embeddings followed by a learned fully-connected layer (i.e., we do not fine tune the entire BERT-based \encoder, just the final output layer). 
We report our results in Figure~\ref{fig:lesion}.
Without end-to-end Transformer fine-tuning, \sys slightly outperforms using just a pretrained encoder (\textsc{-rl}) at times. 
We primarily observe benefits in text-heavy workloads, where the pretrained encoder provides meaningful representations (QA, S, R, EM-T) without relying on positional information from Transformers. 
While we do not perform an exhaustive comparison of pretrained embeddings or non-Transformer-based architectures, this shows that \sys's architecture is a strong out-of-the-box baseline. 

\subsubsection{Remove MLM Pretraining (\textsc{-mlm})}
We eliminate MLM pretraining from the pipeline, and report our results in Figure~\ref{fig:lesion} as \textsc{-mlm}.
\sys meets or exceeds this setting in all but QA, though by a smaller margin than the previous experiments (up to 20.5\% in IMDb-hard). 
MLM pretraining is not effective for QA as the corpus is drawn from Wikipedia---one of the two corpora BERT was trained with---and both base and auxiliary datasets consist of the same, primarily textual vocabulary. 
In such cases, users may wish to disable MLM pretraining, although performance is not significantly impacted.
In contrast, of the non-EM tasks, MLM pretraining is most helpful for FJ IMDb-hard, where random perturbations result in many records with words that are not in the original vocabulary. 

\subsubsection{No Negative Sampling (\textsc{-NS})}
We replace hard negative sampling with random sampling of unrelated records, and report our results in Figure~\ref{fig:lesion}. 
Hard negative sampling tends to improve performance by a similar margin as MLM pretraining, and only negatively impacted the EM-T Company dataset (by up to 8.72\% absolute recall). 
We find larger improvements when the join condition is more ambiguous than recovering a potentially obfuscated join key, as in tasks S (30\% R@1), QA (15\% R@1), and EM-D (15\% R@1).
We use BM25-based sampling for all but FJ and QA, where we have prior knowledge, and develop custom samplers based on Jaccard similarity, and sentence origin, respectively.
This improved performance over the BM25-based sampler by 1\% absolute recall for FJ, and 8.7\% and 5.6\% absolute recall@1 and recall@10, respectively, for QA.

\begin{figure*}
\includegraphics[width=\textwidth]{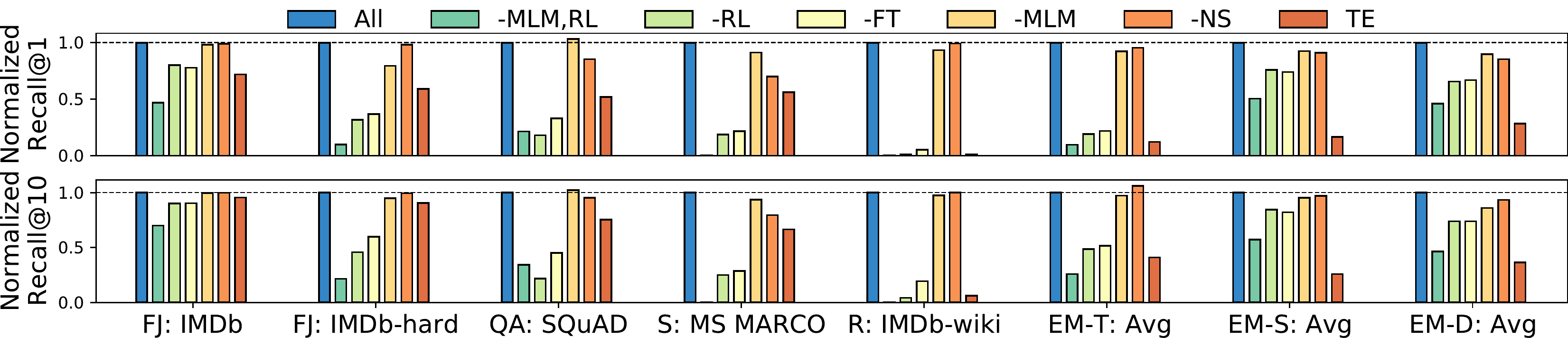}
\caption{Architecture lesion displaying recall@k normalized to the default seen in Table~\ref{tab:baselines} (All) for: a pretrained DistilBERT$_{\text{base}}$ model with no representation learning (\textsc{-mlm,rl}), MLM pretraining with no RL (\textsc{-rl}), RL with no Transformer fine tuning (\textsc{-ft}), RL with no MLM pretraining (\textsc{-mlm}), no hard negative sampling (\textsc{-ns}), and one encoder per dataset (\textsc{te}).}
\label{fig:lesion}
\end{figure*}

\subsubsection{\encoder Configuration (\textsc{te})}
We use two \encoders, one for each dataset, instead of our default single \encoder configuration, and report our results in Figure~\ref{fig:lesion}. 
We find that using two \encoders performs up to two orders of magnitude worse than using a single encoder, especially for strictly structured datasets.
We observe that through the course of the \encoder training procedure, the performance of using two identically initialized \encoders often \emph{degrades}---inspecting the resulting embeddings even when running \sys over two identical tables shows that the exact same terms diverge from one another in the training process.
However, we still provide the option to use a distinct encoder for each data source for future extension to non text-based data, such as audio or image.

\subsubsection{Index Optimization}
For \texttt{INNER} and \texttt{FULL OUTER JOIN}s, we optimize for join execution time by indexing the larger dataset, which reduces the number of queries made by the system.
Due to \encoder training, this reflects a small fraction of the end-to-end pipeline execution time at our data sizes, but can be a substantial cost at scale.
We evaluate this optimization by running the joining step while indexing the larger dataset, and then the smaller dataset for \emph{all} tasks. 
On average, this optimization reduces join time by \red{1.76}$\times$, and up to \red{2.81}$\times$. However, this improvement is only meaningful in the MS MARCO workload, saving \red{2.7} minutes. 

\subsection{Extensibility: End-to-End Workloads}
\label{subsec:extensibility}
We show how to extend and use \sys in an end-to-end context. 

\minihead{Entity Matching} Recent deep-learning-based EM systems focus on the matching phase of that two-part (i.e., blocking and matching) end-to-end EM pipeline~\citep{deeper,deepmatcher,ditto}: given a pair of candidate records, these systems must identify if the pair correspond to the same entity.  
For end-to-end EM, \sys must generate candidate blocks with a low rate of false negatives such that it can be efficiently followed by downstream matchers; we verify this in Table~\ref{table:dm_all} (R@10).
Perhaps surprisingly, we find that treating \sys results from a top-1 query as binary classifier achieves performance that is comparable to or better than previous, custom-built state-of-the-art with respect to F1 score.
We believe incorporating general operators for data augmentation and domain knowledge integration to enable the custom advances presented in the current state-of-the-art EM system, Ditto, may allow \sys to entirely bridge this gap~\citep{ditto}. 

\begin{table}[]
\resizebox{\columnwidth}{!}{%
\begin{tabular}{|c|l|r||r|r|r|r|r|}
\hline
\rowcolor[HTML]{F3F3F3} 
\multicolumn{1}{|c|}{\cellcolor[HTML]{F3F3F3}\textbf{Task}} & \multicolumn{1}{l|}{\cellcolor[HTML]{F3F3F3}\textbf{Dataset}} & \multicolumn{1}{c||}{\cellcolor[HTML]{F3F3F3}\textbf{R@10}} & \multicolumn{1}{c|}{\cellcolor[HTML]{a4c8e9}\textbf{\textsc{emb}}} & \multicolumn{1}{l|}{\cellcolor[HTML]{F3F3F3}\textbf{DM$_\text{RNN}$}} & \multicolumn{1}{c|}{\cellcolor[HTML]{F3F3F3}\textbf{DM$_\text{att}$}} & \multicolumn{1}{c|}{\cellcolor[HTML]{F3F3F3}\textbf{DM$_\text{hyb}$}} & \multicolumn{1}{c|}{\cellcolor[HTML]{F3F3F3}\textbf{Ditto}} \\ \hline
EM-T                                                        & Abt-Buy                                                       & 96.70                                                      & \cellcolor[HTML]{DCEAF7}85.05                            & \cellcolor[HTML]{DCEAF7}39.40                                         & \cellcolor[HTML]{DCEAF7}56.80                                         & \cellcolor[HTML]{DCEAF7}62.80                                         & 89.33                                                       \\ \hline
EM-T                                                        & Company                                                       & 69.89                                                      & \cellcolor[HTML]{DCEAF7}74.31                            & 85.60                                                                 & 89.80                                                                 & 92.70                                                                 & 93.85                                                       \\ \hline
EM-S                                                        & Beer                                                          & 92.86                                                      & \cellcolor[HTML]{DCEAF7}91.58                            & \cellcolor[HTML]{DCEAF7}72.20                                         & \cellcolor[HTML]{DCEAF7}64.00                                         & \cellcolor[HTML]{DCEAF7}78.80                                         & 94.37                                                       \\ \hline
EM-S                                                        & iTunes-Amzn                                                   & 100                                                        & \cellcolor[HTML]{DCEAF7}84.92                            & 88.50                                                                 & \cellcolor[HTML]{DCEAF7}80.80                                         & 91.20                                                                 & 97.06                                                       \\ \hline
EM-S                                                        & Fodors-Zagat                                                  & 100                                                        & \cellcolor[HTML]{DCEAF7}88.76                            & 100                                                                   & \cellcolor[HTML]{DCEAF7}82.10                                         & 100                                                                   & 100                                                         \\ \hline
EM-S                                                        & DBLP-ACM                                                      & 100                                                        & \cellcolor[HTML]{DCEAF7}98.05                            & \cellcolor[HTML]{DCEAF7}98.30                                         & \cellcolor[HTML]{DCEAF7}98.40                                         & \cellcolor[HTML]{DCEAF7}98.45                                         & \cellcolor[HTML]{DCEAF7}98.99                               \\ \hline
EM-S                                                        & Amazon-Google                                                 & 98.94                                                      & \cellcolor[HTML]{DCEAF7}70.43                            & \cellcolor[HTML]{DCEAF7}59.90                                         & \cellcolor[HTML]{DCEAF7}61.10                                         & \cellcolor[HTML]{DCEAF7}70.70                                         & 75.58                                                       \\ \hline
EM-S                                                        & DBLP-Scholar                                                  & 96.29                                                      & 57.88                                                    & 93.00                                                                 & 93.30                                                                 & 94.70                                                                 & 95.60                                                       \\ \hline
EM-S                                                        & Walmart-Amzn                                                  & 94.97                                                      & \cellcolor[HTML]{DCEAF7}69.60                            & \cellcolor[HTML]{DCEAF7}67.60                                         & \cellcolor[HTML]{DCEAF7}50.00                                         & 73.60                                                                 & 86.76                                                       \\ \hline
EM-D                                                        & DBLP-ACM                                                      & 99.96                                                      & \cellcolor[HTML]{DCEAF7}97.58                            & \cellcolor[HTML]{DCEAF7}97.50                                         & \cellcolor[HTML]{DCEAF7}97.40                                         & \cellcolor[HTML]{DCEAF7}98.10                                         & 99.03                                                       \\ \hline
EM-D                                                        & DBLP-Scholar                                                  & 95.99                                                      & 58.08                                                    & 93.00                                                                 & 92.70                                                                 & 93.80                                                                 & 95.75                                                       \\ \hline
EM-D                                                        & iTunes-Amazon                                                 & 100                                                        & \cellcolor[HTML]{DCEAF7}64.65                            & 79.40                                                                 & \cellcolor[HTML]{DCEAF7}63.60                                         & 79.40                                                                 & 95.65                                                       \\ \hline
EM-D                                                        & Walmart-Amzn                                                  & 95.39                                                      & \cellcolor[HTML]{DCEAF7}67.43                            & \cellcolor[HTML]{DCEAF7}39.60                                         & \cellcolor[HTML]{DCEAF7}53.80                                         & \cellcolor[HTML]{DCEAF7}53.80                                         & 85.69                                                       \\ \hline
\end{tabular}%
}
\caption{Recall@10 for \sys (R@10) and F1 Scores for the EM workloads using \sys with \texttt{join size = 1} (\textsc{emb}) compared to deep-learning-based EM solutions from DeepMatcher (DM$_*$) and Ditto. \sys meets or exceeds (highlighted) at least one recent, state-of-the-art classifier in most workloads. \sys has a low false negative rate (1 - R@10), and can be used \textit{with} these methods to increase precision}
\label{table:dm_all}
\vspace{-25pt}
\end{table}

\minihead{Search} In MS MARCO passage ranking, results are ranked in relevance order, and rankings are evaluated via mean reciprocal rank (MRR) of the top 10 results. 
We rank \sys results based on their query distance, and compare MRR@10 with existing MS MARCO  baselines.
We obtain \red{MRR@10 of $0.266$} on the dev set after just 2.5M training examples, outperforming the official Anserini BM25 baseline solution of $0.167$.
Our results exceed K-NRM, a kernel based neural model, with a dev set MRR@10 of 0.218~\citep{knrm}, and is slightly below the state-of-the-art from the time, IRNet, with 0.278 MRR@10~\citep{marcohomepage}. 
For comparison, the first state-of-the-art BERT-based model uses BERT$_\text{large}$ to achieve MRR@10 of 0.365 after training with 12.8M examples~\citep{bertmarco}, and current state-of-the-art is 0.439~\citep{rocketqa}. 
By developing an additional joining operator, \sys can implement ColBERT~\citep{colbert}, a previous state-of-the-art method that achieves 0.384 MRR@10 in MS MARCO document ranking that operates on larger input passages: we must remove the pooling step from \sys's \encoder (1 line config change), and develop a \retriever that indexes and retrieves bags of embeddings for each record. 

\minihead{Recommendation} In Application B of task R, we estimate the IMDb ratings of movies in the test set given a training dataset and Wikipedia corpus.
We report the mean squared error (MSE) between the rating estimates and the true rating.
The task is easy: predicting the average training rating returns 1.19 MSE, and a gradient-boosted decision tree (GBDT) over the joined data returns 0.82 MSE.
We aim to show extensibility while meeting GBDT performance.

There are two approaches to enable this analysis with \sys by reusing the representation learned for the IMDb-wiki task: similarity defined in a single hop in terms of the labeled training data, or by first going through the Wikipedia data in two hops.

In the first method, users configure \sys to index labeled IMDb training data (with ratings) and retrieve records related to the test data (without ratings) via an \texttt{INNER JOIN}. 
Users then post-process the output by averaging the labels of the returned \records.
In the second method, users require two \texttt{LEFT JOIN}s that index the Wikipedia training data against the IMDb test dataset, and the IMDb train dataset against the Wikipedia training data. 
Following a two-hop methodology, users first retrieve the closest Wikipedia summary to each test IMDb record, and then 
retrieve the closest labeled IMDb instances to each previously retrieved Wikipedia summary, which is post-processed via averaging. 

Both approaches reuse the pretrained encoder from the IMDb-wiki task, and require at most \red{6} configuration changes and a 7-line post-processing module.
We report results when aggregating with a \texttt{join size} of 1, 10, 20, and 30; both approaches improve performance as the number of neighbors increase, until a plateau is reached. 
The two-hop approach attains MSE of 1.69, 0.92, 0.89, and 0.89 for \texttt{join size} 1, 10, 20, and 30, respectively, while the one-hop approach performs better with 1.59, 0.89, 0.86, and 0.85.

\subsection{Low Development Effort}
\label{subsec:easy}

We describe \sys's configuration to show the low number of config changes we made, and comment on the supervision required.

\subsubsection{Configuration}
Perhaps surprisingly, we found that exposing only join specification parameters (Listing~\ref{listing:config}) is a strong out-of-the-box baseline.
We require input datasets and supervision to follow a fixed naming convention, under which obtaining the results in Table~\ref{tab:baselines} relies on a default set of configurations options, where only the data directory must be changed.
In Table~\ref{table:datasets}, we list how many config changes are required for the best results from Figure~\ref{fig:lesion}.
\eat{
\begin{lstlisting}[frame=single,caption={Core configuration lines annotated with options.},captionpos=b,float=t,label={listing:config}]
\begin{Verbatim}[commandchars=\\\{\}]
\small
\dred{"data_dir"}: \dred{"IMDb-wiki"},      
\dred{"join_type"}: \dred{"INNER"},    \textcolor{teal}{# or "LEFT", "RIGHT", "FULL"}
\dred{"left_size"}: 1,
\dred{"right_size"}: 1,
\end{Verbatim}
\end{lstlisting}
}


\begin{lstlisting}[frame=single,caption={Core configuration lines annotated with options.},captionpos=b,float=t,label={listing:config}, language=Python,basicstyle=\small,commentstyle=\textcolor{teal},stringstyle=\dred]
"data_dir": "IMDb-wiki",      
"join_type": "INNER", # or "LEFT", "RIGHT", "FULL"
"left_size": 1,
"right_size": 1,
\end{lstlisting}

We generate each result in Figure~\ref{fig:lesion} by altering the following unsurfaced options: number of encoders, encoder intialization, encoder finetuning (boolean), fraction of supervision, negative sampler.
Users can additionally specify lower-level hyperparameters, which we fix across our experiments: epochs, batch size, embedding size, pooling type, tokenizer, learning rate, loss function parameters.

\subsubsection{Dependence on Labeled Data}
A key user input is labeled examples of relevant record pairs.
With the exception of MS MARCO, we use all available labeled examples of \textit{relevant pairs} in Table~\ref{tab:baselines} and Figure~\ref{fig:lesion}.
In MS MARCO, we use 2.5M of the 397M of the provided labeled \textit{triples}, which is just $0.63\%$ of the provided triples.
For the remaining datasets, we evaluate how many labeled examples are required to match performance of our reported results. 
We found that 9 of the 17 datasets required all of the provided examples.
The remaining datasets required $54.4\%$ of labeled relevant data on average to meet Recall@1 performance, and $44.5\%$ for Recall@10. 
In the best case, for EM-S DBLP-ACM, only $30\%$ and $1\%$ of the data is required to achieve the same Recall@1 and Recall@10, respectively.
\section{Related Work}
\label{sec:relwork}

\minihead{Similarity Joins} Similarity-based, or fuzzy, joins often focus on the unsupervised setting~\citep{surajit,fj1,fj2}. 
State-of-the art systems such as AutoFJ~\citep{autofj} are tailored for the case where tables can be joined using exact keys or shared phrases that do not differ greatly in distribution (e.g., our EM tasks)---not complex context enrichment.
\sys generalizes to arbitrary notions of similarity and join semantics at the expense of supervised pairs, and can also leverage unsupervised methods for negative sampling or MLM pretraining.

\minihead{Automated Machine Learning} Automated Machine Learning (AML) systems aim to empower users to develop high-performance ML pipelines minimal intervention or manual tuning. They support various types of data preprocessing, feature engineering, model training and monitoring modules. Examples of AML tools include Ludwig~\citep{ludwig}, Overton~\citep{overton}, Google Cloud AutoML~\citep{gcloud}, and H20.ai~\citep{h20}. However, these platforms do not focus on context enrichment, leaving it as an exercise for users to perform prior to data ingestion. 

\minihead{Relational Data Augmentation}
Relational data augmentation systems seek to find new features for a downstream predictive task by deciding whether or not to perform standard database joins across a base table and several auxiliary tables~\citep{hamlet, hamletpp, arda}. 
Similar to context enrichment, these systems aim to augment a base table for a downstream ML workload. 
However, they assume the existence of KFK relationships that simply must be uncovered. 

\minihead{Data Discovery} Data discovery systems find datasets that may be joinable with or related to a base dataset, and to uncover relationships between different datasets using dataset schema, samples, and metadata~\citep{dd1,dd2,dd3,dd4,dd5,dd6,dd7}.
These systems typically surface KFK relationships, and do not tune for downstream ML workloads.

\minihead{NLP and Data Management} A recurring aim in data management is to issue natural language commands to interface with structured data~\citep{nlidb,csc}. Recent work in Neural Databases~\citep{neuraldb} aims to replace standard databases with Transformer-powered schemaless data stores that are updated and queried with natural language commands. Related to this work are systems that leverage advances in NLP to provide domain-specific functionality, such as converting text to SQL~\citep{text2sql1,text2sql2}, correlating structured and unstructured data~\citep{tabert}, enabling interpretable ML over structured data~\citep{tabnet}, or automating data preparation~\citep{rpt}. We focus on the broader problem of context enrichment of entities for downstream tasks. To our knowledge, this has not been formally addressed in the literature, which often assumes entity information has been resolved~\citep{neuraldb}.
\section{Conclusion}
\label{sec:conclusion}

We demonstrate how seemingly unrelated tasks spanning data integration, search, and recommendation can all be viewed as instantiations of context enrichment.
We propose keyless joins as a unifying abstraction that can power a system for general context enrichment, which allows us to view context enrichment as a data management problem. 
Consequently, we developed and applied \sys, a first-of-its kind system that performs no-code context enrichment via keyless joins. 
We evaluate how developing a keyless join enrichment layer empowers a single system to generalize to five downstream applications, with no ML code written by the user.



\eat{\begin{figure*}
\includegraphics[width=\textwidth]{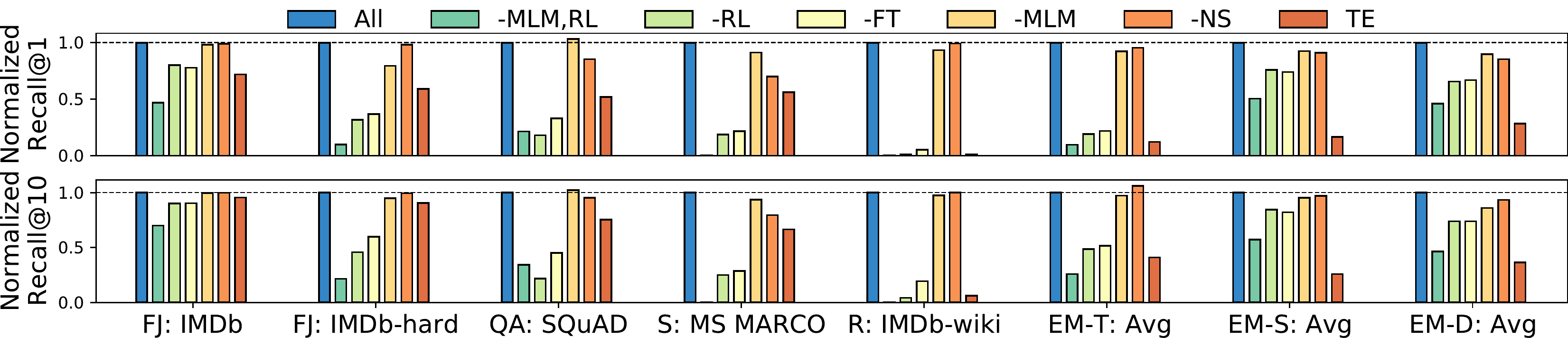}
\caption{alternative to lesion -- what do you think?}
\label{fig:lesion}
\vspace{-1em}
\end{figure*}}


\bibliographystyle{ACM-Reference-Format}
\bibliography{sample}

\end{document}